\documentclass[12pt,preprint]{aastex}

\def\Msun{M$_{\odot}$}

\def\han {\mbox{{\rm H}$\alpha$}}
\def\ha{\han~}

\def\hb {{\rm H}$\beta$}
\def\mic{~$\mu$m}

\def\spose#1{\hbox to 0pt{#1\hss}}
\def\simlt{\mathrel{\spose{\lower 3pt\hbox{$\mathchar"218$}}
     \raise 2.0pt\hbox{$\mathchar"13C$}}}
\def\simgt{\mathrel{\spose{\lower 3pt\hbox{$\mathchar"218$}}
     \raise 2.0pt\hbox{$\mathchar"13E$}}}
\def\lsim{\rlap{$<$}{\lower 1.0ex\hbox{$\sim$}}}
\def\gsim{\rlap{$>$}{\lower 1.0ex\hbox{$\sim$}}}
\def\lam{$\lambda$}

\begin{document}

\pagestyle{myheadings}

\title{Star Formation in Emission-Line Galaxies Between Redshifts of 0.8 and 1.6}

\author{Erin K. S. Hicks\altaffilmark{1}, Matthew A. Malkan\altaffilmark{1}, Harry I. Teplitz\altaffilmark{2}, Patrick J. McCarthy\altaffilmark{3}, Lin Yan\altaffilmark{4}}

\altaffiltext{1}{Department of Physics and Astronomy, University of California,   Los Angeles, CA, 90095-1562}
\altaffiltext{2}{NASA Goddard Space Flight Center, Code 681, Greenbelt, MD 20771}
\altaffiltext{3}{Observatories of the Carnegie Institution of Washington, 813 Santa Barbara Street, Pasadena, CA 91101}
\altaffiltext{4}{SIRTF Science Center, California Institute of Technology, 1200 East California Boulevard, Pasadena, CA 91125}

\begin{abstract}
Optical spectra of 14 emission-line galaxies representative of the 1999 NICMOS parallel grism \ha survey of McCarthy et al. are presented.  Of the 14, 9 have emission lines confirming the redshifts found in the grism survey.  The higher resolution of our optical spectra improves the redshift accuracy by a factor of 5.  The [O II]/\ha values of our sample are found to be more than two times lower than expected from Jansen et al.  This [O II]/\ha ratio discrepancy is most likely explained by additional reddening in our \han-selected sample [on average, as much as an extra E(B-V) = 0.6], as well as to a possible stronger dependence of the [O II]/\ha ratio on galaxy luminosity than is found in local galaxies.  The result is that star formation rates (SFRs) calculated from [O II] \lam3727 emission, uncorrected for extinction, are found to be on average 4 $\pm$ 2 times lower than the SFRs calculated from \ha emission.  Classification of emission-line galaxies as starburst or Seyfert galaxies based on comparison of the ratios [O II]/\hb~and [Ne III] \lam3869/\hb~is discussed.  New Seyfert 1 diagnostics using the \ha line luminosity, {\it H}-band absolute magnitude, and \ha equivalent widths are also presented.  One galaxy is classified as a Seyfert 1 based on its broad emission lines, implying a comoving number density for Seyfert 1s of $2.5^{+5.9}_{-2.1}~\times$ 10$^{-5}$ Mpc$^{-3}$.  This commoving number density is a factor of 2.4$^{+5.5}_{-2.0}$ times higher than estimated by other surveys. 
\end{abstract}


\section{Introduction}

Recent studies have found that the comoving integrated star formation rate (SFR) increases by an order of magnitude as z increases from 0 to 1 (Connolly et al. 1997; Hogg et al. 1998, hereafter H98; Yan et al. 1999).  Surveys at redshifts greater than z = 2 show that this trend flattens or decreases in that range, indicating that there may be a peak in the SFR within the redshift range $1 < z < 2$.  Thus, it is necessary to observe galaxies at these redshifts to understand star formation history not only in individual galaxies, but also in the universe as a whole. 

SFRs for local galaxies have been inferred by measuring the \ha \lam6563 emission line (Kennicutt 1983).  \ha is a good SFR indicator because it measures the flux of ionizing photons from young, massive O and B stars.  Assuming an initial mass function (IMF) and case B recombination, it is then possible to calculate the total SFR.  This method is observationally easiest up to z $\sim$ 0.3, beyond which \ha is redshifted out of the optical wavelength range.  Other prominent optical emission lines are \hb~and [O III] \lam5007, which are observable up to z = 0.7, as well as [O II] \lam3727 and [Ne III] \lam3869, which are still in the optical up to a redshift of z = 1.2.  It is possible to use [O II] \lam3727 (a blend of [O II] \lam\lam~3726,3729) as a SFR indicator, although it is not as reliable (Kennicutt 1992).  Recent studies of local galaxies (Jansen, Franx, \& Fabricant 2001; Carter et al. 2001; Charlot et al. 2002) have found that the observed L([O II])/L(\han) ratio is correlated with the galaxy luminosity.  These studies have shown that the dependence is due to both reddening and the metallicity-dependent excitation of the interstellar medium.  The SFRs derived from the [O II] fluxes therefore have large uncertainty.  In local galaxies Jansen et al. (2001) found this uncertainty to be a factor of $\sim$ 3, if no correction for metallicity and dust is possible.  Although this wide range in [O II]/\ha values will lead to large uncertainty in SFRs based on [O II] line flux, it is still valuable to determine an average value for calibration of [O II] and SFR.  Using this average value will not give accurate SFRs for individual objects; but, for a large survey, accurate SFR densities may still be attainable, making it possible to assess the SFR in the previously inaccessible redshift range of $1 < z < 2$.  

McCarthy et al. (1999, hereafter M99) used parallel grism observations with NICMOS onboard the Hubble Space Telescope (HST) to survey blank fields for \han-emitting galaxies at $0.7 < z < 1.9$.  The survey, using the G141 slitless grism on NICMOS, covered approximately 64 arcmin$^{2}$ and found 33 emission line galaxies.  Redshifts were measured with the assumption that the single observed emission line was \han.  \ha is the most likely identification, since \ha is the strongest optical/near-IR line (Kennicutt 1992).  M99 found an average SFR per galaxy of 30 M$_{\odot}$ yr$^{-1}$ and an emission-line galaxy comoving number density of $3.3 \times 10^{-4}~ h^{3}_{50}~ Mpc^{-3}$.  This comoving number density is about half that of present-day galaxies brighter than L* in the {\it B}-band (Ellis et al. 1996) and similar to bright Lyman break galaxies at z $\sim$ 3 (Steidel et al. 1996).  The calculated SFR density (Yan et al. 1999) is consistent with results from a similar \ha study by Hopkins, Connolly, \& Szalay (2000).  Two of the 33 M99 galaxies were suggested to be active galaxies (Seyferts) based on their high equivalent widths and compact morphologies.

To confirm the redshifts given by M99, and thus the SFRs and comoving number density estimates, we have obtained optical spectra of 14 of the galaxies in the M99 survey.  Along with the redshifts of the galaxies, the SFRs based on the [O II] emission are considered in an effort to determine if [O II], instead of \han, can be used for future surveys.  The comoving number density of Seyfert 1 galaxies is also investigated.  Section 2 describes the observations, reduction process, and discusses how well our 14 galaxies represent the 33 galaxies in the survey done by M99, and $\S$3 presents the emission line detections.  The implications of these emission lines (SFR, reddening, classification of the galaxies as active or starburst, and Seyfert 1 luminosity function), as well as two new diagnostic diagrams, are given in $\S$4.

Throughout this paper, $H_{0} = 50~ km~ s^{-1}~ Mpc^{-1}$ and $q_{0} = 0.5$ are adopted, unless stated otherwise, to simplify comparison with previous studies.

\section{Observations and Data Reduction}
All observations were made with the Keck II telescope using the Low Resolution Imaging Spectrograph (LRIS; Oke et al. 1995, Cohen et al. 1994).  Three gratings were used: the 150 line mm$^{-1}$ grating, which is blazed at a wavelength of 7500 \AA; the 300 line mm$^{-1}$, which is blazed at a wavelength of 5000 \AA; and the 400 line mm$^{-1}$, which is blazed at a wavelength of 8500 \AA.  The dispersions are 4.80 \AA~pixel$^{-1}$, 2.48 \AA~pixel$^{-1}$, and 1.90 \AA~pixel$^{-1}$, for the 150, 300, and 400 line mm$^{-1}$ gratings respectively.  A slit width of 1$^{\prime\prime}$ was used for all observations, which was usually comparable to the seeing full-width half-maximum (FWHM).  The FWHM of sky emission lines were measured to be 20.0 \AA~for the 150 line mm$^{-1}$ grating, 14.4 \AA~for the 300 line mm$^{-1}$ grating, and 12.0A~for the 400 line mm$^{-1}$ grating, giving respective widths of 860 km s$^{-1}$, 620 km s$^{-1}$, and 510 km s$^{-1}$ at 7000 \AA.  At the blaze wavelength the resolution is R $\sim$ 375, 350, 700, for the 150, 300, and 400 line mm$^{-1}$ gratings, respectively.  All observed emission lines, except where noted otherwise, were spectrally unresolved.  Table 1 lists the length of exposure, number of exposures, filter and grating used, grating angle, and the rest wavelengths observed for each galaxy.

\subsection{Data Reduction}
Reduction of the spectra followed standard techniques, using the CCDRED and LONGSLIT packages in the IRAF software package\footnote[5]{IRAF is
  distributed by NOAO, which is operated by AURA Inc., under cooperative agreement with the NSF.}.  The reduction began with removal of cosmic rays, which was done with the COSMICRAYS package.   Bias subtraction was, in most cases, accomplished by subtraction of images nodded along the slit, which subtracts the CCD bias as well as a first order removal of sky emission lines.  The images were subsequently flat fielded and bad pixels were replaced by the average value of the surrounding pixels.  Dark current was found to be negligible (less than 24 electrons hour$^{-1}$).


The IRAF package APALL was used to extract one-dimensional spectra.  Sky subtraction regions were placed between 15 and 30 pixels (3$''$.2 to 6$''$.5) from the center of the galaxy spectrum.  The sky emission lines are independently variable in time and it is often difficult to remove the sky emission completely despite the method used by APALL and the nodded image subtraction.  In some cases this can distort an emission line's profile, which is manifested as a change in line flux and/or strongly non-Gaussian profile.  Sky subtraction is the dominant source of error in the measurements of the galaxy emission lines.  For those galaxies with two exposures, an average of the one-dimensional spectra was taken.


The one-dimensional spectra were then wavelength-calibrated using an arc lamp (neon or neon with argon) emission spectrum.  At least 20 emission lines were used to fit a Chebyshev polynomial to determine the wavelength scale.  For a few of the observations, a calibration frame was not taken in a matching grating setting, and the sky emission lines in the image itself were used instead of an arc.  The fits to the calibration emission lines had a standard deviation of 3.0 pixels, 2.0 pixels, and 1.4 pixels for the 150, 300, and 400 line mm$^{-1}$ gratings, respectively.  This results in a redshift accurate to around 0.004 for the 150 line mm$^{-1}$, 0.001 for the 300 line mm$^{-1}$, and 0.0007 for the 400 line mm$^{-1}$ gratings.  However, the redshift accuracy is decreased because of effects from flexure and, in some cases, from the significant time between acquisition of the galaxy and arc spectra.  In addition, the faint lines, such as those found in the galaxy spectra, are not located as accurately as bright arc lines.  As a result, the redshifts are only accurate to 0.002 for both the 300 and 400 line mm$^{-1}$ gratings.  This accuracy is still 5 times better than that of M99.  The grating and grating angles were chosen for each object so that [O II] would fall within the observed wavelength region, assuming the redshift given by M99, as well as avoiding strong OH sky emission.  However, for the July 17 run only the 300 line mm$^{-1}$ grating was used, which lead to spectra without the region expected to contain [O II] for some galaxies.  For the higher redshift objects in our sample, it was not possible to obtain a spectrum of the region expected to contain [O II] with any configuration.  The only exception is the spectrum obtained of J0055+8515a, for which the 150 line mm$^{-1}$ grating was available.  The range of rest-frame wavelengths covered, assuming the redshift given by M99, are listed in Table 1.


For flux calibration of the spectra we used Oke (1990) spectrophotometric standards, observed and reduced via the above method.  For all but five of the objects, the spectrophotometric observations used for calibration were taken during the same run as the target observation.  During the July 1998 and January 1999 runs, however, no spectrophotometric standards were observed.  Standards observed during the May 1999 run were used for calibrating this data.  Comparing the transmission curve of a spectrophotometric standard observed during the December 1999 run with one from the May 1999 run reveals a difference of less than 3\%.  Therefore, the transmission curve does not change significantly over extended lengths of time, and using a standard observed several months earlier or later for calibration does not add much additional error to the flux.  There was also no spectrophotometric standard observed in August 1998, when J0055+8515a was observed using the 150 line mm$^{-1}$ grating.  The observations from the May 1999 run, using the 400 line mm$^{-1}$ grating, were assumed to have a flux proportional to the flux that would be observed if the 150 line mm$^{-1}$ grating were used.  This assumption is reasonable since the blaze wavelengths and grating angles are similar.  The ratio of flat fields taken with each grating gives the constant of proportionality needed to calibrate J0055+8515a using a standard observed using the 400 line mm$^{-1}$ grating.  It was found that the 400 line mm$^{-1}$ grating has a flux 2.67 times lower per pixel than what would be measured if using the 150 line mm$^{-1}$ grating at 7000 \AA, with a slight decrease (increase) at higher (lower) wavelengths.  Using the May 1999 standard for the calibration of J0055+8515a does not add more than an additional 10\% of error.

The overall uncertainty in the flux calibration is typically about 8\%.  Including the error due to the placement of the continuum in measuring line fluxes and the noise of the spectrum, this error increases to around 19\% (37\% in regions of poor sky subtraction).  Another factor in determining the total line fluxes of the entire galaxy is loss of light due to the slit used during the spectroscopic observations.  A comparison of the continuum flux measured in six of the spectra and the flux measured in direct images reveals an average slit loss of 2.76$\pm$0.37 times the flux.  Individually measured slit loss corrections were used when possible, and the average slit loss was used for those galaxies without a direct measurement.  A slit loss correction has been applied to all line fluxes (see Table 3 for the slit loss correction used [SLC]).  For those galaxies with direct slit loss measurements, the error introduced in the resulting line fluxes is that of the continuum {\it R}-band spectroscopic measurement and direct {\it R}-band image measurement, typically a total of 3\%.  The additional error in the line fluxes of those galaxies without a direct measurement is simply the standard deviation of the average slit loss correction that was applied, which is 13\%.  This gives a total error in the line fluxes of 20\% up to 39\%, depending on the quality of the sky subtraction and the slit loss correction used.  As discussed above, there is also additional error in the flux of the July 17 1998, August 1998, and January 1999 data due to the use of a spectrophotometric standard observed during a run other than that of the galaxy observation.  The spectra have not been corrected for Galactic extinction.

Imaging photometry was obtained in the Johnson {\it R} band (\lam$_{c}$=6417 \AA, FWHM=1185 \AA) and in some cases the Johnson {\it I} band (\lam$_{c}$=8331 \AA, FWHM=3131 \AA).  The direct CCD images were bias-subtracted and then divided by a normalized flat field.  The multiple exposures of a given field were then combined.  Table 2 lists the details of the exposures taken for each field: date of exposure, combined exposure length, and the filter used.  Rather than {\it R}-band images, {\it I}-band images were taken for J0040+8505a and J0055+8518a and both {\it I}- and {\it R}-band images were available for the J1237+6219 field.  No image was obtained of J0622-0018a.  Photometry was measured in a circular aperture of diameter 10 pixels (2$''$.2), which is about twice the FWHM of the sources.  To calibrate the magnitudes, photometry of standard stars at similar airmasses were taken using the same radius aperture.  All observing runs had good photometric conditions and the consistency of the standards is 2\%.  As a result, the magnitude error is dominated by a combination of the photon statistics of the standard star and galaxy measurement.  The {\it H}-band magnitudes and {\it J-H} colors given by M99 are listed in Table 3, using the Vega scale, for each of the galaxies in our study.  Also included are the colors measured in our study, as well as the error on these measurements. 


J0613+4752a has a higher uncertainty in its magnitude because it is less than 2$''$ from a star.  This causes contamination of the galaxy at our resolution.  To measure the magnitude of J0613+4752a, photometry in a circular 3 pixel radius aperture was measured for the galaxy and a corresponding region on the opposite side of the star as well.  Assuming the star has a symmetric light profile, the light from the star on the opposite side of the galaxy can then be subtracted from the measurement made centered on the galaxy.  This yields the light contribution from just the galaxy, which is reported in Table 3.

\subsection{Representation of the \ha Grism Survey}
Our sample of galaxies is a fair representation of the galaxies in the 33 galaxy survey discovered by M99, with a slight bias towards the lower redshifts.  Figures 1, 2, and 3 show the distribution of apparent H magnitude, redshift (as given by M99), and \ha luminosity, respectively, of the spectroscopic targets chosen for our sample, as well as the entire M99 sample.  A Kolmogorov-Smirnov test reveals that the probability that the two distributions, the full sample and our subsample, differ is 0.79, 0.31, and 0.37, for the apparent H magnitude, redshift, and \ha luminosity distributions, respectively.  Thus, the difference in the distributions is not statistically significant.  The observation of J0040+8505a was taken during the end of the observing run, and the twilight was too bright for an adequate spectrum of the faint galaxy to be obtained.  However, if strong features, such as those typical of a Seyfert 1, were present in J0040+8505a, they should still have been detected.  As discussed, the spectrum of galaxy J0613+4752a is contaminated by a nearby star, most likely preventing any emission lines from being detected.  Therefore, the results of this study, excluding J0040+8505a and J0613+4752a, can be generalized to all the objects in the M99 \ha survey.

\section{Results}
\subsection{Detected Emission Lines}
For most of the galaxies, [O II] (the strongest emission line after \ha in typical starburst galaxies) was expected to fall within the observed wavelength range, based on the redshifts published by M99.  Other typically strong starburst emission lines expected (based on local surveys) to be seen in our optical/near-UV are, in approximate order of decreasing relative strength, H$\gamma$, [Ne III] \lam3869, C II] \lam2326, C III] \lam1909, [Ne IV] \lam2424, [O III] \lam4363, and Mg II \lam2798 (McQuade, Calzetti, \& Kinney 1995; Kennicutt 1992).  Of the fourteen galaxies observed, nine had detectable emission lines, and all of those confirmed the redshift given by M99.  [O II] was observed in each of the galaxies for which a redshift was obtained (with the exception of J1237+6219c for which the wavelength range did not include [O II]) along with at least one other line, most often Mg II or [Ne III].  For seven of the galaxies, three or more emission features were observed at redshifts consistent with M99.

If the galaxy is not a starburst, but rather an active galaxy, its spectrum could differ.  For example, a Seyfert 1 is identified by its broad (V $= 10^{3}$ to $10^{4}$ km s$^{-1}$) permitted emission lines, e.g., the Balmer series and Mg II.  The identification of a Seyfert 2 or a low ionization nuclear emission-line region (LINER) requires more effort, which will be discussed in $\S$4.6. 

\subsection{Discussion of Individual Objects}
Of the 14 objects for which spectra were obtained, redshifts for nine of the objects confirm M99's results.  Of these nine galaxies, one is a Seyfert 1, while the others exhibit typical starburst emission lines.  Figure 4 shows the regions of spectra containing emission lines for each galaxy.  A more detailed plot of the region of the spectrum containing the [O II] emission is shown in Figure 5.  Table 4 lists emission features observed, their known rest wavelengths, observed wavelengths, flux, significance levels, total error in the flux, and equivalent widths, along with the redshift we determine.  The total error in the line fluxes not only includes the error due to noise in the spectrum, but also the error due to the various flux calibration steps discussed in $\S$2.1.  In each case the redshift measured confirms that given by M99, with z$_{us}-$ z$_{M99}$ = 0.001 $\pm$ 0.006.  We were unable to measure any emission features in the spectra of the remaining five objects.  Table 5 contains upper limits on emission lines for these unconfirmed galaxies, under the assumption that the redshift based on \ha is correct.  Two of the unconfirmed galaxies, J0040+8505a and J0613+4752a, did not have adequate spectra, for reasons discussed below; however upper limits are still given in Table 5.  The other unconfirmed galaxies had no strong emission lines to allow a redshift determination.

There are three possible explanations for a lack of detected emission lines:  (1) the galaxies could be heavily reddened, which would significantly decrease the flux of the continuum and/or lines in the observed wavelength region (optical/near-UV); (2) the line observed by M99 is not \han, and we are therefore looking at a part of the spectrum devoid of strong emission lines; or (3) the emission features were just too weak to be detected.  The average color of our galaxy sample is R-H $\sim$ 2.6, indicating the continua of the galaxies are not heavily reddened.   Galaxies with a confirmed redshift have an average R-H $\sim$ 2.9 compared to R-H $\sim$ 1.9 for unconfirmed galaxies.  Excluding J0613+4752a, which has an uncertain magnitude measurement, and J0931-0049a, which is expected to be bluer because it is a Seyfert 1, the average for all galaxies changes to R-H $\sim$ 2.8.  Among this sample, confirmed galaxies have an average R-H $\sim$ 3.0 and unconfirmed galaxies R-H $\sim$ 2.3.  Since the unconfirmed galaxies are not redder than our confirmed galaxies, it is possible to eliminate the hypothesis that continuum reddening alone is the reason some galaxies were unable to have redshifts confirmed.  The reddening of the galaxies in this sample is discussed further in $\S$4.2.  Should the lack of features in our spectra be due to a misidentification of the line detected by M99, other possible identifications are [O III] \lam\lam5007, 4959, H$\beta$~\lam4861, and [O II] \lam3727, observed at even higher redshifts than M99 supposed.  The possibility of one of these other lines being a more likely identification of the feature detected by M99 is discussed further for the individual unconfirmed galaxies.  All galaxies were unresolved, and their spatial profiles along the slit are consistent with a point source.  Notes on the individual galaxies follow.

{\it\#1, J0040+8505a -} This galaxy was observed at the very end of an observing night.  As a result, the twilight dominates the spectrum, eliminating the possibility of detecting weak features that may be present in the galaxy spectrum.  Upper limits are reported, although they are not expected to place very accurate constraints on the emission line fluxes, since the detected flux is mostly due to twilight.  However, if strong emission features were present, like those typical of Seyfert 1s (Osterbrock 1989), then these features would have been detected.  Therefore, based on its upper limits, this galaxy is not a Seyfert 1.

{\it\#2, J0050+8518a} - This emission-line galaxy has six $\geq$ 3 $\sigma$ emission lines and five weaker features, which confirmed the redshift to be 0.758 $\pm$ 0.003.  The overall high strength of the lines could suggest that the galaxy is a Seyfert 2 (McCarthy 1993).  However, the relatively weak [Ne III]/\ha compared to [O II]/\ha (see $\S$4.6) leads us to classify the galaxy as a strong starburst.

{\it\#3, J0613+4752a} - No believable emission features are present in the spectrum of this object.  The proximity of this galaxy (about 2$''$) to an $R = 16.6$ mag star causes starlight contamination in the spectrum of this galaxy.  The majority of the flux detected in the spectrum is starlight; thus the upper limits reported for this object do not place very tight constraints on the actual galaxy line flux.  According to the average broad line strengths of Seyfert 1s given by Osterbrock (1989), this object cannot be rejected, based on its upper limits, as a possible Seyfert 1.  Assuming the redshift published by M99, z = 1.04, [O II] should be located at 7603 $\pm$ 56 \AA~which is where the A band atmospheric absorption is located.  This indicates that if strong [O II] emission were present, then it might have been missed due to this atmospheric feature.  It is not possible to determine if one of the alternative identification discussed above for the emission line detected by M99 is more likely than another.

{\it\#4, J0622-0018a} - This emission-line galaxy has three $\geq$ 3 $\sigma$ emission lines and five weaker features, which confirmed the redshift to be 1.135 $\pm$ 0.002.  The strongest line, [O II], is in a region of strong sky emission.  The sky subtraction was not perfect and resulted in altering the shape (and thus the flux measurement) and possibly the central wavelength of the feature.  This causes a redshift estimate based on [O II] to be only 1.123, which deviates from 1.135 by more than our estimated error in redshift.  For this reason, the [O II] feature is not included in the redshift estimate.  Based on the expected observed wavelength of [O II] and instrumental FWHM, it can be estimated that as much as 15\% of the flux may have been missed because of poor sky subtraction.  The overall high strength of the lines could suggest that the galaxy is a Seyfert 2 (McCarthy 1993).  However, the relatively weak [Ne III]/\ha compared to [O II]/\ha (see $\S$4.6) leads us to classify the galaxy as a strong starburst. 

{\it\#5, J0741+6515a} - Three emission features are detected in the spectrum of this galaxy: [O II] \lam2470, Mg II, and [O II] \lam3727.  The redshift is confirmed to be z = 1.445 $\pm$ 0.003.  [Ne III], which, if present, would be located in the observed spectrum, is not seen, and an upper limit is reported.

{\it\#6, J0741+6515b} - There are no believable emission features in the spectrum obtained of this galaxy.  C III] would be the strongest expected feature at the wavelengths observed, with [O II] not available.  The upper limits on these lines eliminate the possibility of this object being a Seyfert 1 (Osterbrock 1989).  It is unlikely that the feature detected by M99 is something other than \ha because the other possibilities discussed above imply redshifts above 2.7.  The signal-to-noise of the continuum is about half of that in the spectra of confirmed galaxies.

{\it\#7, J0741+6515c} - Three emission features are measured in this spectrum: [O II], H$\delta$, and H$\gamma$.  The redshift is found to be z=1.061 $\pm$ 0.004, confirming the redshift given by M99.   The [O II] line is within two pixels (3.8 \AA) of the peak of a relatively strong sky emission feature, and because of imperfect sky subtraction, its measurement is uncertain.  Based on the expected observed wavelength and expected FWHM, it is estimated that at most 8\% of the [O II] line flux may have been missed.  [Ne III], H$\epsilon$, and [O III] \lam4363 are not detected, and upper limits are given.

{\it\#8, J0931-0449a} - Two observed broad emission lines, C III] (6441 km s$^{-1}$) and Mg II (3809 km s$^{-1}$), demonstrate that the galaxy is a Seyfert 1.  Two additional emission features are observed, giving a confirmed redshift of z = 0.978 $\pm$ 0.005.  A weak [O II]/\ha value of 0.03, typical of Seyfert 1s (Osterbrock 1977), is also present.  The two broad lines show a spatial profile consistent with a point source along the slit (see Figure 6), indicating no spatial structure on the scales observed.  The full spectrum is shown in Figure 7.  The R-H color of this galaxy, R-H = 1.8, also confirms it is a Seyfert since it is 0.8 magnitudes bluer than the average of the other galaxies in this sample.  Its J-H color, given by M99, is also one of the bluest in the sample.

{\it\#9, J1039+4145a} - There are two detections for this galaxy, [O II] \lam2470 and [O II] \lam3727.  This gives a redshift of z = 1.481 $\pm$ 0.001, confirming the redshift based on \han.  [Ne III] may also be present, although the central wavelength of the feature is off by more than the expected error.  It is, therefore, not included in the final redshift estimate.  

{\it\#10, J1120+2323a} - Both Mg II and [O II] are observed; however, unexpectedly for a starburst (McQuade et al. 1995), the [O II] line is the weaker of the two.  The [O II] line is in a region of strong sky emission, and because of imperfect sky subtraction, its measurement is uncertain.  Based on the expected observed wavelength and expected FWHM, it is estimated that at most 10\% of the [O II] line flux may have been missed.  H$\epsilon$~and [S II] \lam4072 are also present.  The redshift given by the above lines, z = 1.383 $\pm$ 0.004, is consistent with the published redshift of 1.37 based on \han.

{\it\#11, J1120+2323b} -  [O II] is observed, along with H$\gamma$~and [Ne III].  Again, [O II] falls in a part of the spectrum with strong sky emission, possibly causing the emission to be affected.  At most 10\% of the [O II] line flux may have been missed, based on the expected observed wavelength and FWHM.  The three emission lines give a redshift of z = 1.135 $\pm$ 0.004, confirming the redshift of 1.13 based on \han.

{\it\#12, J1134+0406a} - No emission features are detected in the spectrum of this galaxy.  For the redshift determined by M99, [O II] should be located at 7156 $\pm$ 56 \AA.  This wavelength is not in a region with strong sky emission.  If the feature detected by M99 was not \han, but instead [O III] or H$\beta$, then the wavelength region observed in this study would not be expected to contain strong emission features.  The strongest features would be, in decreasing strength, C II] \lam2326, C III] \lam 1909, [Ne IV] \lam2424, and Mg II.  Mg II was detected in four of our seven confirmed galaxies for which it was available.  If instead the M99 feature were [O II] then the redshift would be 2.38 and Ly$\alpha$ would be expected to be present.  Thus, it is possible that this galaxy's redshift was not confirmed because the part of the spectrum observed is not expected to contain strong emission features and the M99 feature is either [O III] or H$\beta$.  It is equally plausible that reddening prevented a confirmation.  The upper limits given in Table 5, which assume the feature measured by M99 is \han, eliminate the possibility of this object being a Seyfert 1 (Osterbrock 1989).

{\it\#13, J1237+6219a} - No emission features are detected in this spectrum.  At the wavelengths observed, [O II] is not available.  Ly$\alpha$ and C III] should be available, although neither is present.  Based on the upper limits of these features, this object is not a Seyfert 1 (Osterbrock 1989).  It is not possible to determine if the M99 infrared line has been misidentified.  

{\it\#14, J1237+6219c} - The spectrum of this galaxy has two detected lines with $\ge$ 2$\sigma$: C IV \lam1550 and C III] \lam1909.  Mg II is also detected with a 1.1 $\sigma$ feature.  These features give a redshift of z = 1.638 $\pm$ 0.003, confirming the redshift given by M99, z = 1.64.  Ly$\alpha$ and [O II] \lam2470 are also within the wavelength region observed, although neither is seen and upper limits are reported.

\section{Discussion}
\subsection{Intrinsic [O II]/\ha Ratio}
A wide range in the [O II]/\ha in local galaxies has been noted by several authors (see, e.g., Hammer et al. 1997).  This range is partly attributed to a dependence of [O II]/\ha on galaxy luminosity, with higher ratios for less luminous galaxies (see, e.g., Jansen et al. 2001).  The observed [O II]/\ha ratios for our galaxies are given in Table 6.  For 4 of the14 galaxies, our spectra did not include the wavelength where [O II] was expected to be located; thus no [O II]/\ha is measured for these objects.  Three of these four spectra are those with no spectral features detected and thus no redshift determined.  The fourth is J1237+6219c, for which the redshift given by M99 was confirmed on the basis of three features at wavelengths shorter than [O II].  Upper limits are given in Table 6 for the two galaxies, J0613+4752a and J1134+0406a, for which the expected location of [O II] is observed but no feature detected.  For those galaxies in our sample for which [O II] was observed, the average [O II]/\ha is 0.18 $\pm$ 0.12, if J0931-0449a, a Seyfert 1, and upper limits are excluded.  However, the average of the logarithmic ratio for these objects is log([O II]/\han) = -0.82 $\pm$ 0.27, or [O II]/\ha = 0.15$^{+0.13}_{-0.07}$.  Furthermore, if we had instead calculated the median value so that the significant upper limit of J1134+0406a would be included, then the median would drop to log([O II]/\han) = -0.97, or [O II]/\ha = 0.11.

Our observed [O II]/\ha is significantly less than the median ratio observed by Kennicutt (1992), which was [O II]/\ha = 0.45 $\pm$ 0.26.  Figure 8 shows a histogram of the observed [O II]/(\han+[N II]) ratios in the Kennicutt sample, as well as our sample for comparison.  A value for the [N II]/\ha ratio must be assumed for lower resolution spectra, such as some of the Kennicutt sample spectra and all of the M99 sample.  For those spectra in the Kennicutt sample observed at higher resolution, the mean observed [N II]/\ha = 0.5.  However, as stated by Kennicutt, this value may be slightly biased toward higher values, because galaxies with blended lines were not included, which led to excluding objects with lower [N II]/\ha values.  Gallego et al. (1997) observed an average [N II]/\ha = 0.4.  If this lower value is assumed, then the mean observed value of Kennicutt's sample decreases to [O II]/\ha = 0.42 $\pm$ 0.24.  Using this same lower [N II]/\ha value decreases our average observed [O II]/\ha value to 0.17 $\pm$ 0.11.  This dependence of [O II]/\ha on the assumed [N II]/\ha ratio, which could result in a difference of at least 7\%, must be kept in mind when considering the results of further calculations.

The average {\it B}-band absolute magnitude, M$_{B}$, of the Kennicutt sample differs by more than half a magnitude compared to our sample; therefore, a difference in the [O II]/\ha [and thus the [O II]/(\han+[N II]) ratio] is expected.  The anti-correlation between galaxy luminosity and [O II]/\ha in local galaxies (Jansen et al. 2001) predicts a difference in the observed [O II]/\ha values for the two samples of 11\%.  The relationship used is that given by Jansen et al. (2001) for [O II]/\ha, uncorrected for interstellar reddening and using $H_{0} = 50~ km~ s^{-1}~ Mpc^{-1}$.  Table 6 lists M$_{B}$ for each galaxy in our sample, as well as the corresponding predicted log([O II]/\han) value according to the Jansen et al. relationship.  The predicted [O II]/\ha value for the Kennicutt sample is 0.53$^{+0.17}_{-0.13}$, based on an average M$_{B}$ = -20.80 $\pm$ 1.4.  For our sample of 14 galaxies, the average M$_{B}$ is -21.33 $\pm$ 1.25, which gives a predicted [O II]/\ha = 0.49$^{+0.13}_{-0.11}$.  For starburst galaxies with confirmed redshifts, the average M$_{B}$ is -20.99 $\pm$ 0.88 resulting in a predicted [O II]/\ha = 0.52$^{+0.10}_{-0.08}$.  

In Figure 9, it can be seen that the majority of our galaxies, seven out of nine, have significantly lower [O II]/\ha than predicted by the Jansen et al. curve.  The remaining two objects have values consistent with the Jansen et al. prediction; however, one object only has an upper limit.  J0622-0118a is placed on the plot even though it lacks a second broad band measurement necessary to calculate an accurate M$_{B}$ (indicated by its large error bars).  M$_{B}$ was calculated using the {\it H}-band measurment given by M99 and assuming {\it R-H} = 2.8, the average for our sample (see $\S$ 3.2).  However, J0622-0118a is expected to be consistent with the Jansen prediction, given its high [O II]/\ha of 0.40, which places it on the plot at log([O II]/\han) = -0.39.  The other galaxies in our sample lack [O II] measurements, for reasons discussed in the $\S$ 3.2, and are thus not included in Figure 9.

Glazebrook et al. (1999) obtained [O II]/\ha for 13 galaxies with z $\sim$ 1.  If only galaxies with L(\han) greater than 1$\sigma$ are considered (eight galaxies total), their average [O II]/\ha is 0.67 $\pm$ 0.33, or log([O II]/\han) = -0.17$^{+0.17}_{-0.29}$.  At higher redshifts, z $\sim$ 3, Pettini et al. (2001) found an average [O II]/H$\beta$ = 2.0 $\pm$ 0.9 for five galaxies.  If \han/H$\beta$ = 6 is assumed, this gives an average [O II]/\ha = 0.33 $\pm$ 0.14, or log([O II]/\han) = -0.48$^{+0.15}_{-0.24}$.  Including these two studies in Figure 9, it can be seen that both Glazebrook et al. and Pettini et al. agree with the relationship given by Jansen et al.  Assuming any Balmer decrement in the plausible range of 3 - 10 to convert the Pettini et al. data from [O II]/H$\beta$ to [O II]/\ha results in consistency with the Jansen et al. relationship.  

  
Possible explanations for our lower [O II]/\han, compared to those studies
mentioned above, are additional reddening and/or an intrinsically lower [O
II]/\han.  As Jansen et al. (2001) demonstrated, metallicity and reddening
are equally responsible for the observed range in [O II]/\ha in local
galaxies.  Both of these properties are correlated with the luminosity of
the galaxy, leading to lower intrinsic [O II]/\han, as well as more
reddening in higher luminosity galaxies.  As a result, more luminous
galaxies have significantly lower values of [O II]/\ha than less luminous
ones.  As Figure 9 demonstrates, the range in [O II]/\ha seen in our
sample cannot be fully explained by the dependence of [O II]/\ha on
luminosity found in the Jansen et al. sample of local galaxies.
Arag\'{o}n-Salamanca et al. (2002) found a stronger dependence of [O
II]/\ha on galaxy luminosity in the Universidad Complutense de Madrid
(UCM) sample of local \han-selected galaxies (see Fig. 9).  The mean
observed [O II]/\ha in their sample is half that observed by Jansen et al.
Arag\'{o}n-Salamanca et al. attribute this discrepancy between their
sample and that of Jansen et al. to a difference in average extinction
because of the selection technique used (\ha versus {\it B}-band
selection). Our sample of \ha selected galaxies is more likely to have an
[O II]/\ha dependence closer to that found by Arag\'{o}n-Salamanca et al.,
and thus our lower observed [O II]/\ha can be partially explained.  Using
the [O II]/\han-M$_{B}$ relationship found by Arag\'{o}n-Salamanca et al.,
our predicted average [O II]/\ha value for confirmed starburst galaxies is
[O II]/\ha = 0.23$^{+0.06}_{-0.05}$, which, within the errors, is
consistent with our observed average [O II]/\ha = 0.18 $\pm$ 0.12.
However, as mentioned, if the average of the logarithmic ratios or
the median is considered, then our observed [O II]/\ha is significantly
below their prediction.  Therefore, the [O II]/\han-M$_{B}$ relationship
at higher redshifts cannot be assumed to be equivalent to that observed
locally.  As suggested by Hammer et al. (1997), reddening and metallicity
might play an even more important role in the observed [O II]/\ha at high
redshifts.  If a wider range of metallicity and/or reddening is present at
higher redshifts than in local galaxies, then a wider range of observed [O
II]/\ha is realistic.  A more detailed discussion on the explanation for
our lower observed [O II]/\ha follows in $\S$ 4.5.


\subsection{Additional Reddening}
Assuming an intrinsic [O II]/\ha value of 0.52 from the Jansen et al. (2001) relationship, it is possible to estimate how much additional reddening might be present in our galaxies.  The averaged interstellar extinction curve used throughout this section is that given by Seaton (1979), but adopting A$_{V}$/E(B-V) = 3.09, instead of 3.20.  For our sample, the average confirmed starburst galaxy extinction required is E(B-V)$_{[O II]/\han}$ = 0.59 $\pm$ 0.34 mag.  The required extinction for each individual galaxy is given in Table 6.  If our galaxies have an intrinsically lower [O II]/\ha than predicted by Jansen et al., then this calculation results in an overestimate of the actual extinction.  


Another measure of the extinction in each galaxy is obtainable from the observed Balmer emission line ratios.  However, the Balmer lines detected have uncertain fluxes due to unknown stellar absorption lines underlying the emission.  Thus, the resulting E(B-V)$_{Balmer}$ is an upper limit. Using the \ha emission given by M99 in combination with any H$\gamma$, H$\delta$, or H$\epsilon$ emission detected in this study, the E(B-V)$_{Balmer}$ of each galaxy can be found (see Table 6).  The averaged interstellar extinction curve and intrinsic Balmer line ratios used are those given by Osterbrock (1989).  For those galaxies with multiple Balmer lines detected, an average E(B-V)$_{Balmer}$ is given.  For some galaxies, only upper limits on the higher Balmer line flux were available.  In these cases a lower limit to the E(B-V)$_{Balmer}$ is measured, meaning the upper limit on the extinction is not lower than the reported E(B-V)$_{Balmer}$ in Table 6.  The average extinction for our sample, using this method, is E(B-V)$_{Balmer}$ = 0.63 $\pm$ 0.53 magnitudes.  


\subsection{Star Formation Rate Based on [O II]}
\ha is a reliable indicator of the current SFR (Kennicutt 1983), using the calibration
\begin{equation}
SFR(M_{\odot}~yr^{-1}) = 8.9 \times 10^{-42}L(\han)E(\han),
\end{equation}
\noindent where E(\han) is the extinction at \ha \lam6563.  This can be transformed into a calibration with [O II] using the average observed relationships [O II]/(\han+[N II]) = 0.30 and accounting for [N II] \lam\lam6583, 6548, which is blended with \ha in low resolution spectra, by using [N II]/\ha = 0.5 (Kennicutt 1992).  These values give the observed value of [O II]/\ha = 0.45, which then results in the following relationship
\begin{equation}
SFR(M_{\odot}~yr^{-1}) = 2.0 \times 10^{-41}L([O II])E(\han).
\end{equation}
\noindent The [O II]/\ha ratio is the observed value for nearby galaxies (Kennicutt 1992), uncorrected for reddening.  Therefore, no further extinction correction at [O II] is needed, if it is assumed that the reddening in our sample of galaxies is the same as in nearby galaxies.  

Table 6 lists the luminosity of the [O II] emission for each galaxy in our sample along with the resultant SFRs using equation (2) and no extinction [E(\han) = 0 magnitudes = 1 times the flux].  Table 6 also contains the SFRs based on \ha with the [N II] correction used above applied.  Therefore, the \ha SFRs in our table are 33\% lower than those given by M99 in their Table 3, which did not correct for [N II].  Since J0931-0449a is a Seyfert 1, the line emission is not due to high-mass stars.  Thus, the SFR for J0931-0449a is an upper limit, but it is included in the table for completeness.  Some of the spectra did not reach far enough into the blue to measure [O II] emission, and thus no SFRs were measured for these galaxies.  After correcting for [N II], M99 found an average SFR of 21 $M_{\odot}~yr^{-1}$.  Assuming no extinction at \han, we find that, for confirmed starburst galaxies (all of our confirmed galaxies except J0931-0449a, which is a Seyfert 1), the average SFR based on our measured [O II] is 3 $\pm$ 2 times lower than that given by the \ha measured in M99.  

If an extinction of E(\han) = 1 mag, which corresponds to E(B-V) = 0.43 mag, typical of nearby spiral galaxies (Kennicutt 1983), is assumed, then the [O II] and \ha relationships change to:
\begin{equation}
SFR(M_{\odot}~yr^{-1}) = 2  \times 10^{-41}L(\han)
\end{equation}
and
\begin{equation}
SFR(M_{\odot}~yr^{-1}) = 5 \times 10^{-41}L([O II]).
\end{equation}
\noindent The resulting SFRs based on [O II] are listed in Table 6 in column (11).  Since this extinction is applied to both equations (1) and (2) resulting in equations (3) and (4), respectively, the ratio of SFR(\han)/SFR([O II]) remains unchanged when this extinction is included.  Thus, the average SFR([O II]) is still 3 $\pm$ 2 times lower than SFR(\han).  

The above relationships do not take into account the possible dependence of [O II]/\ha on luminosity (Jansen et al. 2001).  Since the average luminosity of our sample is slightly less than that of the Kennicutt sample (see $\S$4.1), the intrinsic [O II]/\ha is expected to be different, and thus the relationships given in equations (2) and (4) are not quite accurate for our sample.  If an intrinsic [O II]/\ha of 0.52, which is the prediction based on the Jansen et al. relationship for our averaged confirmed starburst galaxies, is used, then the new conversion from L([O II]) to SFR is
\begin{equation}
SFR(M_{\odot}~yr^{-1}) = 4 \times 10^{-41}L([O II]).
\end{equation}
\noindent  The SFRs for each galaxy, using their predicted [O II]/\ha to derive the SFR-L([O II]) relationship, are given in column (12) of Table 6.  Making this correction for the [O II]/\ha dependence on luminosity, it is found that the averaged SFR is 4 $\pm$ 2 times lower than that found by M99.  It can therefore be concluded that our sample of galaxies has an intrinsically lower [O II]/\ha than that predicted by the Jansen et al. (2001) relationship and/or they are reddened more than the E(\han) = 1 mag found in local galaxies by Kennicutt (1992).

Including the average extinction calculated in $\S$4.2, which is based on the [O II]/\ha ratio, results in SFR relationships that account for the maximum amount of additional reddening that may be present in our sample.  Adding this additional reddening to the E(\han) of 1 mag already included by Kennicutt in the SFR relationships, gives a total average reddening of E(B-V) = 1.02 $\pm$ 0.34 mag.  This larger reddening, which we consider more realistic, increases the intrinsic \ha and [O II] fluxes as follows:  
\begin{equation}
L(\han)_{int} = 3.52^{+3.72}_{-1.81}  \times L(\han)_{obs}
\end{equation}
and
\begin{equation}
L([O II])_{int} = 12.59^{+41.34}_{-9.65} \times L([O II])_{obs}.
\end{equation}
The transformations to SFR for are then
\begin{equation}
SFR(M_{\odot}~yr^{-1}) = 8 \times 10^{-41}L(\han)
\end{equation}
and
\begin{equation}
SFR(M_{\odot}~yr^{-1}) = 50 \times 10^{-41}L([O II]).
\end{equation}
The SFR for each galaxy, using its calculated additional reddening, is given in column (13) of Table 6.  Although including this additional reddening has the effect of raising the SFRs based on [O II], so that SFR(\han) equals SFR([O II]), it also has the effect of raising the SFRs based on \ha.  As is seen in the next section, this would result in an SFR density that is higher than reported by any other study.  



\subsection {Star Formation Rate Density}
Assuming our sample is representative of the whole \ha survey, we arrive at an uncorrected, volume-averaged SFR 4 $\pm$ 2 times lower than that found by Yan et al. (1999), which is based on the M99 \ha survey.  Their result was a volume-averaged SFR at z=1.3 $\pm$ 0.5 of 0.13 \Msun~yr$^{-1}$~Mpc$^{-3}$.  Therefore, we find, based on observed [O II] emission with no additional correction for reddening, a volume-averaged SFR of 0.03$^{+0.03}_{-0.02}$ \Msun~yr$^{-1}$~Mpc$^{-3}$ in the same redshift range.

Comparing our volume-average SFR to other studies in a similar redshift range, we find that our [O II]-based SFR density is again lower.  In addition to using [O II] emission as a SFR indicator, \ha emission (Gallego et al. 1995; Tresse \& Maddox 1998; Glazebrook et al. 1999; Yan et al. 1999; Hopkins, Connolly, \& Szalay 2000) and UV continuum (Lilly et al. 1996; Connolly et al. 1997; Treyer et al. 1998) have also been used.  Although both of these methods agree that there is an order of magnitude increase in the SFR from z = 0 to z = 1, the UV estimates are significantly less than those based on \ha emission.  Figure 10 shows our volume-averaged SFR density, along with results from the studies mentioned above.  It can be seen that our results are most consistent with SFR densities based on the UV continuum.  Our SFR density is at least 3 times lower than any results found using the \ha emission method.  The systematic discrepancy between SFR based on the UV continuum and \ha emission can most likely be explained by dust extinction (see, e.g., Steidel et al. 1999), which suppresses the UV continuum more than the \ha emission.  This same reddening may be responsible for the difference in the SFR predictions based on \ha and [O II].

Hogg et al. (1998, hereafter H98) measured [O II] in 375 faint z $\la$ 1 galaxy spectra, making it possible for our results to be compared to a study based on the same SFR indicator.  They found an order of magnitude increase in the SFR density from present day out to z $\sim$ 1, similar to results based on the UV continuum and \han.  Their volume-averaged SFR agrees with results based on \ha (see Fig. 10), possibly indicating that the [O II] to SFR conversion in equation (4), which assumes reddening typical of local spiral galaxies, is correct for z $\la$ 1 galaxies.  Their redshift range is a bit lower than ours, but there is enough overlap for a reasonable comparison.  Despite the use of the same SFR indicator, our SFR density from [O II] is almost 3 times lower than H98's results in a similar redshift range.  

Since \ha was not measured in the H98 sample of galaxies, the intrinsic [O II]/\ha is unknown.  However, it is reasonable to assume that the intrinsic value is near [O II]/\ha $\sim$ 0.45, since using the SFR-L([O II]) relationship given by Kennicutt (1992) results in SFR densities similar to those found in other studies using Kennicutt's relationship for SFR-L(\han).  Another way to determine [O II]/\ha is to use the Jansen et al. relationship.  For the H98 sample, which has an average M$_{B}$ = -20.5$^{+1.4}_{-1.9}$, the relationship predicts an intrinsic [O II]/\han = $0.58^{+0.19} _{-0.18}$ (see Fig. 9).  M$_{B}$ was calculated based on the average {\it R}-band magnitude of the H98 sample, {\it R} = 21.8 $\pm$ 0.8, the average redshift, z = 0.65 $\pm$ 0.2, assuming an average {\it R-H} = 2.8 (the average of our sample; see $\S$ 3.2).  Changing the assumed {\it R-H} color to 0.5 or 5, changes [O II]/\ha by less than 5$\%$.  If the dependence of [O II]/\ha on luminosity is ignored, then it is concluded that our sample has a SFR density at least 2.9 times lower that that found by H98.  Even after accounting for the differences in luminosity, our sample has a SFR density 2.6 times lower than the H98 sample.  The remaining discrepancy between the samples is most likely due to a difference in the amount of reddening.  It is also possible that the dependence of [O II]/\ha on luminosity is greater than that predicted by the Jansen et al. relationship at higher redshifts, which could also play a role in the remaining discrepancy.  Why such a difference would exist in such similar samples is discussed in the next section. 

\subsection{Luminosity Function Comparison}
Comparing our study and H98 in more detail, we calculate the luminosity function (LF) predicted from our [O II] measurements.  This can be done by assuming our sample is representative of the whole M99 sample, and transforming the \ha LF given in Yan et al. (1999), which is based on the M99 \ha galaxies, into a [O II] LF.  The transformation is applied to the \ha LF given in Figure 1 of Yan et al. by assuming a constant [O II]/\ha = 0.18 $\pm$ 0.12 for all galaxies, which is our average observed value.  The new LF then represents what the same 33 galaxy sample would yield if they were all observed in [O II].  Comparing our LF in Figure 11a to the LF given by H98 (only considering their $0.35<z<1.5$ galaxies), our LF is lower, especially at the higher luminosities.  The \ha LF found by Hopkins et al. (2000), transformed using the same constant [O II]/\ha value, is also included in Figure 11a.  The transformed Hopkins et al. luminosity function also appears to confirm
that using our observed [O II]/\ha ratio to transform from an \ha to an [O
II] luminosity function results in a luminosity function that is lower
than that of H98 at the highest luminosities.  However, the Hopkins et al.
luminosity function may be incomplete at the higher luminosities.  The
discrepancy between the Hopkins et al. and Yan et al. luminosity functions
is most likely explained by Yan et al.'s much wider areal coverage, which
allowed them to sample the higher end of the luminosity function more
completely.  Even accounting for incompleteness of the Hopkins et al.
sample at the higher luminosities, our observed [O II]/\ha still results
in a transformed luminosity function that is significantly lower than the
H98 luminosity function at the higher luminosities.

The discrepancy between the transformed \ha LFs and the H98 [O II] LF may be due to the use of a single value for the \ha to [O II] transformation for a sample that we know has a wide range in [O II]/\ha values.  To account for the wide range in [O II]/\ha observed in this study, the value of [O II]/\ha used in the transformation must take into account the value's dependence on the \ha luminosity.  This \ha luminosity dependent [O II]/\ha would then be applied to the \ha LF, resulting in a more accurately transformed LF.  However, the seven galaxies in our sample for which [O II]/\ha were measured, only cover a very narrow range in log(L(\han)) values, 41.9 to 42.2.  This, combined with the error on each individual measurement, leads to a fit with very large uncertainty.  The resulting transformed \ha LFs have such large error bars that no conclusion can be made about the consistency of the transformed \ha and [O II] LFs.  

An alternative to transforming the \ha LFs is to do the transformation in the other direction, i.e., transforming the H98 [O II] LF into a \ha LF, which can then be compared to the Yan et al. (1999) and Hopkins et al. (2000) results.  Our sample of seven galaxies covers a wider range in L([O II]), enabling a more reliable fit to be obtained.  It is assumed that a straight line is a good approximation of the dependence of [O II]/\ha on L([O II]).  The seven data points appear linear to the eye, and more complicated functions could not be justified with so little data.  The linear Pearson's correlation coefficient for a straight line fit is 0.92.  Since our sample is very small, the resulting LF is still uncertain.  In addition, the differences in the SFR density based on [O II] between our sample and the H98 sample may indicate that [O II]/\ha dependence on L([O II]) may be different for the two samples.  Thus, using a transformation based on our sample causes even greater uncertainty in the resulting transformed [O II] LF.  The [O II]/\ha values used for the eight H98 points, from left to right in Figure 11b, are as follows: 0.10, 0.10, 0.10, 0.16, 0.36, 0.52, 0.70, 0.90.  For the first three points, [O II]/\ha=0.10 was assumed, instead of the negative value given by the fit.  As Figure 11b demonstrates, using these more accurate values results in the H98 transformed [O II] LF being slightly lower than the LFs given by Yan et al. and Hopkins et al. at the highest luminosities, although they agree within the errors.

The H98 [O II] LF is higher than the \ha LFs in Figure 11a and slightly lower in Figure 11b as a result of our observed [O II] values being lower than those in the H98 study.  As previously stated, the local luminosity-[O II]/\ha relationship, is not enough to explain the difference in the LFs between our sample and that of H98.  If the dependence of [O II]/\ha on luminosity is stronger in higher redshift galaxies, then the LF discrepancy may be expected.  Also, the differences at the bright end of the LF could be due to the small numbers.  With the steep slope in the \ha LF, small systematic errors could lead to large differences in the derived \ha LF.   

Should neither of the above scenarios be responsible for the discrepancy in the LFs, the difference is most likely attributable to additional reddening in our sample.  This does not necessarily mean the two surveys observed galaxies from completely separate populations.  It is possible that the two studies are observing galaxies from a single population of galaxies, but that the subsamples are from different sides of a broad distribution.  Thus, the differing selection effects imposed on the galaxy population, which has an intrinsically wide range of [O II]/Ha values, causes the two samples to contain different cuts of the same population.  The M99 sample was selected on \ha luminosity in a grism survey.  This method has a bias towards bright, compact sources, which causes our sample to have a higher average absolute {\it B}-band luminosity than those in the H98 sample, which were selected by {\it R}-band magnitude.  Our higher-luminosity objects lead to a lower intrinsic [O II]/\ha, and thus galaxies with lower [O II] are more likely to be observed in our sample.  As stated, this difference in intrinsic [O II]/\ha alone does not, most likely, explain the difference in LFs of the two samples, unless the dependence on luminosity is greater at higher redshifts than observed in local galaxies.  To explain the large discrepancy in LFs, it is also necessary to assume that our galaxies have more reddening (see $\S$4.2).  Based on the {\it R}-band magnitudes of our 14 galaxies, only four would have been selected by H98, suggesting that, on average, the M99 galaxies have more dust extinction than the H98 sample.  Thus, if a sample is chosen based on \ha emission, lower [O II]/\ha than those observed in an {\it R}-band magnitude selected sample are more likely to be observed due to the increased reddening and lower intrinsic [O II]/\ha ratios.  The selection methods for the two samples may, therefore, have limited the surveys to different subsections of the total population, causing neither one to sample the entire population completely.  

Further support for this theory is provided by recent work by
Arag\'{o}n-Salamanca et al. (2002).  As mentioned in $\S$ 4.1, they found
that their \han-selected sample of local galaxies had significantly more
reddening than the local Jansen et al. (2001) sample, which was {\it
B}-band-selected.  This additional reddening resulting in an observed mean
[O II]/\ha that is half that observed in the Jansen et al. sample.
Therefore, even at low redshift, there is such a wide range in the
distribution of [O II]/\ha and E(B-V) that different selection methods
select subsamples with significantly different properties.

This same argument can be used to explain why the Pettini et al. (2001) study, which consists of z $\sim$ 3 galaxies, has [O II]/\ha values consistent with the Jansen et al. relationship, while we do not.  The Pettini et al. sample has an average {\it R-H} = 1.2 $\pm$ 0.2, if {\it H-K} = 1.2 is assumed, which is a reasonable estimate for z $\sim$ 3 galaxies (Chen et al. 2002).  This {\it R-H} color is much lower than our average, found to be {\it R-H}=2.6$\pm$1.0.  The {\it H-K} color must be assumed to be less than 0.8 for the average color of the Pettini et al. sample to be consistent with our average color.  Therefore, it is likely that our sample has more reddening then the Pettini et al. sample, thus leading to a lower observed [O II]/\ha.  However, the  Glazebrook et al. sample, which is also consistent with the Jansen et al. relationship with a sample at z $\sim$ 1, cannot be explained in this way.  The {\it J-H} colors for their sample are similar to our sample, indicating that there is no significant difference in the degree of reddening.  The Glazebrook sample was selected based on {\it I}-band magnitude, and if {\it R-I} = 2.0 is assumed then only 3 of our galaxies out of the 14 would have been missed in their sample.  The average M$_{B}$ of the two samples only differ by 0.4 mag, and the average redshift of the Glazebrook sample is lower than that of our sample.  However, there is enough overlap in the redshift ranges covered in our sample and the Glazebrook sample, that their average observed [O II]/\ha is not expected to be a factor of 3.7 discrepant.  Although continuum reddening does not seem to be responsible for this discrepancy, it is possible that the difference in the studies is due to reddening in the line emission, instead of the continuum.  

Thus, when deriving the SFR-L([O II]) relationship, it is necessary not only to consider the dependence of [O II]/\ha on luminosity, but also, at least when working at z $\gtrsim$ 1, to assess the degree of reddening.  Because of the wide range in [O II]/\ha in our study and the others mentioned above, it is possible that the E(B-V) extinction in the observed galaxy population can vary by more than a factor of 4 mag.

\subsection{Classifying Emission-Line Galaxies}
Standard diagnostic diagrams (Baldwin, Philips, \& Terlevich 1981) are valuable for classifying emission-line galaxies based on their optical lines.  However, these diagrams are only useful for those galaxies at low enough redshifts for the necessary emission lines to be available at optical wavelengths.  The only diagnostic emission lines generally available at the rest wavelength ranges observed for our eight galaxies are [O II] and [Ne III] \lam3869.  By comparing the luminosity of each of these lines to \hb, it is still possible to classify emission galaxies as either a Seyfert 2 or a starburst galaxy (Rola, Terlevich, \& Terlevich 1997).  This diagnostic does not work well for identification of LINERs.  Using an assumed Balmer decrement of 6, the \ha luminosities published by M99 can be converted into \hb~luminosities.  The contribution of [N II]\lam\lam~6583, 6548 to the \ha + [N II] blend is assumed to be [N II]/\ha $\sim$~0.5 (Kennicutt 1992) for starbursts.  Figure 12 shows the position of our galaxies that have both [O II] and [Ne III] measurements compared to those galaxies observed by Rola et al. (1997).  

All of our galaxies lie within the range for starburst galaxies, which is given by Rola et al. (1997) to be at log([Ne III] \lam3869/\hb) $\la$ -0.2.  Increasing the Balmer decrement would push them closer to the area containing the Seyfert 2s, but still well within the region of starburst galaxies.  An implausibly high decrement, greater than \han/\hb~= 15, is needed in order to classify J0622-0018a, the most likely candidate in our sample, as a Seyfert 2.  Another adjustment that can be made is the contribution of [N II], which is closer to [N II]/\ha $\sim$~1.0 (McQuade et al. 1995) for Seyfert 2s.  This would also increase the ratio [Ne III] \lam3869/\hb, but by a smaller amount than from increasing the Balmer decrement.  Even if a higher contribution of [N II] is used, which can be as high as [N II]/\ha $\sim$~1.5 (see, e.g., Malkan $\&$ Oke 1983), in combination with a reasonably high Balmer decrement, it is still not possible to classify any of our galaxies as a Seyfert 2 using this diagnostic.  We thus conclude that even though J0622-0018a and J0055+8518a have emission features that might lead one to suspect they are Seyfert 2s, they should still be classified as starburst galaxies.

\subsection{New Diagnostic Diagram}
In an effort to identify Seyfert 1s that were assumed to be starburst galaxies in the M99 \ha survey, a new diagnostic diagram is used.  Comparing the rest equivalent width to the luminosity of the \ha line suggests that Seyfert 1s may be separated from starburst galaxies using the criteria that they have $log(L_{\han}) > 42.5$ and $EW(\han) > 100 \AA$ (see Fig. 13).  These limits were determined by eye, guided by the number of Seyfert 1s expected in the M99 sample (see $\S$4.8), and should only be taken to be approximate values.  A separation can also be made comparing the absolute H (1.65\mic) magnitude to the luminosity of \han.  Figure 14 shows that Seyfert 1s may be separated from starburst galaxies by again having $log(L_{\han}) > 42.5$ as well as $M_{H} < -25$.  Galaxies that fit both of the above criteria are most likely Seyfert 1s.  These criteria indicate that not only is J0931-0449a a Seyfert 1, but also that J0917+8142a and J0923+8149a are good Seyfert 1 candidates.  In addition, J0917+8142c is a good candidate due to, as M99 noted, its large \ha equivalent width and compact morphology.  J0917+8142c meets the first criterion, but no {\it H}-band magnitude was obtained by M99 to enable its placement on the $log(L_{\han})$ versus $M_{H}$ plot. It is most likely that it would be place well within the Seyfert 1 area of the plot.  J0738+0507a, although it only satisfies one of the above criterions, is also a good candidate due to its compact morphology observed by M99.

\subsection{Seyfert 1 Number Density}
Of our 14 galaxies, we were able to confirm the redshifts of nine, and of these, one is a Seyfert 1 (J0931-0449a).  Assuming our galaxies are representative of the \ha survey and that a Seyfert 1 could not have been missed in our other eight spectra, this gives a Seyfert 1 comoving number density of $3.7^{+8.4}_{-3.1} \times 10^{-5}~h^{3}_{50}~Mpc^{-3}$; this comoving number density is 1/9 of the total galaxy comoving number density found by M99.  Errors were estimated from Poisson statistics (Gehrels 1986).  If this comoving number density is correct, there should be 2 or 3 other Seyfert 1s in the 33-galaxy \ha survey.  Furthermore, assuming the redshifts of the four unconfirmed galaxies (excluding J0613+4752a which can not be excluded as a possible Seyfert 1; see $\S$3.2) are correct, based on the \ha emission line, and that emission lines would have been detected if they were Seyfert 1s, the comoving number density decreases to 1/13 of the total density found by M99, giving $2.5^{+5.9}_{-2.1} \times 10^{-5}~h^{3}_{50}~Mpc^{-3}$.  In this case there should then be 1 or 2 other Seyfert 1s among the \ha survey galaxies.  If instead it is assumed that J0613+4752a is a Seyfert 1 as well, then the commoving number density is 2/14 of the total density found by M99, giving $4.7^{+10.8}_{-3.9} \times 10^{-5}~h^{3}_{50}~Mpc^{-3}$.  A method to identify these likely Seyfert 1 galaxies was discussed in $\S$4.7, as were the most likely candidates.

The number of quasars/Seyfert 1s of a given luminosity at a given redshift is fairly uncertain at the redshifts of the fourteen galaxies in this study.  Grazian et al. (2000), using a luminosity dependent luminosity evolution (LDLE) model, suggest that the number density of Seyfert 1s within the absolute blue magnitude range of -21 to -25 and redshifts $1.0 < z < 1.4$ is 1.1$^{+0.58}_{-0.36} \times 10^{-5}~Mpc^{-3}$.  We therefore had an expectation value of 0.39 Seyfert 1 galaxies in our sample of 13 galaxies (which excludes J0613+4752a for reasons stated earlier).  The probability of the observation of one Seyfert 1 galaxy in our sample being consistent with the expected value of 0.39 is 26\%.  The consistency of our result with the prediction of the LDLE model cannot, therefore, be ruled out at a high significance level.  However, if we take our observation of one Seyfert 1 galaxy in our sample as representative of the whole M99 sample, then this gives a comoving number density a factor of 2.4$^{+9.6}_{-2.2}$ times higher, which, within the error, is still marginally consistent with the LDLE model prediction.  


\section{Conclusions}
Observations have been made of 14 galaxies, which constitute a subset of the 33 galaxies detected by M99 in an \ha grism survey covering 64 arcmin$^{2}$.  The redshifts of 9 of these 14 galaxies confirmed M99's results, with improved accuracy.  Of the five galaxies with unconfirmed redshifts, two, J0040+8505a and J0613+4752a, did not have adequate spectra.  Either the other three were either too faint (possibly because of reddening), or the emission detected in the \ha survey was not actually \han, and we were looking at a part of the galaxy's spectrum that contains no detectable emission lines.  

Our sample of galaxies have lower observed [O II]/\ha than both the Kennicutt (1992) and Jansen et al. (2001) samples.  These lower [O II]/\ha ratios are most likely due to a greater dependence of the ratio on luminosity at higher redshifts, as well as additional reddening, as compared to local galaxies.  Should the lower [O II]/\ha be due solely to reddening, then our sample has an average E(B-V) = 0.59 $\pm$ 0.34 mag of additional reddening, which, when combined with the E(\han) = 1 mag already included by Kennicutt, results in an average total reddening of E(B-V) = 1.08 $\pm$ 0.34 mag.  

The lower observed [O II]/\ha found in our sample result in a lower volume-averaged SFR density, with our sample an average of 4 $\pm$ 2 times lower than that given by M99.  Comparing our sample to H98, which is also based on [O II], it is found that our volume-averaged SFR density is at least a factor of 2.6 times lower.  This discrepancy between studies based on the same SFR indicator is most likely due to a sample of galaxies with a wide range in [O II]/\ha, this range being caused by a wide range in reddening, as well as intrinsic [O II]/\ha values. 

One galaxy, J0931-0449a, is a Seyfert 1, based on its broad emission lines.  The remaining galaxies are considered starburst galaxies, based on their placement in a diagnostic diagram using [O II] \lam3727, [Ne III] \lam3869, and \hb.  Two new diagnostics were found to be useful in distinguishing between Seyfert 1s and starburst galaxies, using the equivalent width of \han, the luminosity of the \ha emission, and the absolute magnitude in the {\it H}-band.  Using these new diagnostics we predict J0917+8142a, J0917+8142c, J0923+8149a, and probably J0738+0507a, which were not observed in this study, are also likely Seyfert 1 candidates.

It can be assumed that no other Seyfert 1 galaxies are present in our spectroscopic sample (since we would have detected their broad emission lines) and that the 13 galaxies (excluding J0613+8508, which cannot be ruled out as a Seyfert 1 based on upper limits) are representative of the 33 galaxies in the \ha survey.  The probability that detecting one Seyfert 1 in our sample is consistent with the 0.39 expectation value is 26\%.  Should our detection of one Seyfert 1 galaxy out of 13 be representative of the whole M99 sample, then the comoving number density of Seyfert 1 galaxies is $2.5^{+5.9}_{-2.1} \times 10^{-5}~h^{3}_{50}~Mpc^{-3}$, which is 2.4$^{+9.6}_{-2.2}$ times higher than predicted by published LDLE Seyfert 1/quasar LFs.  Within the error, this is consistent with the LDLE model prediction.

This research was supported, in part, by a grant from the Space Telescope Science Institute GO-7499.01-96A.

\references
\reference{} Arag\'{o}n-Salamanca, A., Alonso-Herrero, A., Gil de Paz, A.,
Gallego, J., Zamorano, J., P\'{e}rez-Gonz\'{a}lez, P. G.,
Garc\'{i}a-Dab\'{o}, C. E. 2002, preprint.
\reference{} Baldwin, J., Philips, M., \& Terlevich, R. 1981, PASP, 93, 5
\reference{} Carter, B. J., Fabricant, D. G., Geller, M. J., Kurtz, M. J., \& McLean, B. 2001, ApJ, 559, 606
\reference{} Charlot, S., Kauffmann, G., Longhetti, M., Tresse, L., White, S. D. M., Maddox, S. J., \& Fall, S. M. 2002, MNRAS, 330, 876
\reference{} Chen, H.-W., et al. 2002, ApJ, 570, 54
\reference{} Cohen, J. G., Cromer, J., \& Southard, S., Jr. 1994 ADASS, 3, 469
\reference{} Connolly, A. J., Szalay, A. S., Dickinson, M., Subbarao, M. U., \& Brunner, R. J. 1997, ApJ, 486, 11L
\reference{} Ellis, R., Colless, M., Broadhurst, T., Heyl, J., \& Glazebrook, K. 1996, MNRAS 280, 235
\reference{} Gallego, J., Zamorano, J., Rego, M., \& Vitores, A. G. 1997, ApJ, 475, 502
\reference{} Gallego, J., Zamorano, J., Aragon-Salamanca, A., \& Rego, M. 1995, ApJ, 459, L1
\reference{} Gehrels, N. 1986, ApJ, 303, 336
\reference{} Galzebrook, K., Blake, C., Economou, F., Lilly, S., \& Colless, M. 1999, MNRAS, 306, 843
\reference{} Grazian, A., Cristiani, S., D'Odorico, V., Omizzolo, A., Pizzella, A. 2000, AJ, 119, 2540
\reference{} Hammer, F., et al. 1997, ApJ, 481, 49
\reference {}Hogg, D. W., Cohen, J. G., Blandfor, R., \& Pahre, M. A. 1998, ApJ, 504, 622
\reference{} Hopkins, A. M., Connolly, A. J., \& Szalay, A. S. 2000, ApJ, 120, 2843
\reference{} Jansen, R. A., Franx, M., \& Fabricant, D. 2001, ApJ, 551, 825

\reference{} Kennicutt, R. C. 1992, ApJ, 388, 310
\reference{} Kennicutt, R. C. 1983, ApJ, 272, 54
\reference{} Lilly, S. J., Le Fevre, O., Hammer, F., \& Crampton, D. 1996, ApJ, 460, L1
\reference{} Malkan, M. A., \& Oke, J. B. 1983, ApJ, 265, 92
\reference{} McCarthy, P. J., et al. 1999, ApJ, 520, 548
\reference{} McCarthy, P. J. 1993, ARAA, 31, 639
\reference{} McQuade, K., Calzetti, D., \& Kinney, A. L. 1995, ApJ, 97, 331
\reference{} Okay, J.B. 1990, AJ, 99, 1621
\reference{} Oke, J. B., et al. 1995, PASP, 107, 375
\reference{} Osterbrock, D. E. 1977, ApJ, 215, 733
\reference{} Osterbrock, D. E. 1989, {\it Astrophysics of Gaseous Nebulae and Active Galactic Nuclei}, (Mill Valley: Univ. Science Books)
\reference{} Pettini, M., Shapley, A. E., Steidel, C. C., Cuby, J., Dickinson, M., Moorwood, A. F. M., Adelberger, K. L., \& Giavalisco, M., 2001, ApJ, 554, 981
\reference{} Rola, C. S., Terlevich, E., \& Terlevich, R. J. 1997, MNRAS, 289, 419
\reference{} Seaton, M. J. 1979, MNRAS, 187, 73S
\reference{} Steidel, C., Giavalisco, M., Pettini, M., Dickinson, M., \& Adelberger, K. 1996, ApJ, 462, L17
\reference{} Steidel, C., Adelberger, K. L., Giavalisco, M., Dickinson, M., \& Pettini, M. 1999, ApJ, 519, 1
\reference{} Tresse, L., \& Maddox, S. 1998, ApJ, 495, 691
\reference{} Treyer, M. A., Ellis, R. S., Milliard, G., Donas, J., \& Bridges, T. J. 1998, MNRAS, 300, 303
\reference{}Yan, L., McCarthy, P. J., Freudling, W., Teplitz, H. I., Malumuth, E. M., Weymann, R. J., \& Malkan, M. A. 1999, ApJ, 519, 47L

\clearpage

\begin{deluxetable}{rrlrrlrrrr}																	
\rotate																	
\scriptsize																	
\tablenum{1}																	
\tablecolumns{10}																	
\tablewidth{0pc}																	
\tablecaption{Galaxy Obervations}																	
\tablehead{																	
\cline{1-10} \\																	
\colhead{ID Number} & \colhead{Field} & \colhead{Object}& \colhead{$\alpha$(2000.0)\tablenotemark{a}}&\colhead{$\delta$(2000.0)\tablenotemark{a}}& \colhead{Date} & \colhead{Exposure\tablenotemark{b}} & \colhead{Grating} & \colhead{Grating Angle} & \colhead{Rest \lam Range} \\
\colhead{} & \colhead{} & \colhead{}& \colhead{}&\colhead{}& \colhead{} & \colhead{(s)} & \colhead{(line mm$^{-1}$)} & \colhead{} & \colhead{(\AA)}}																	
\startdata																	

1 & J0040+8505	&	a	&	00 40 43.93	&	85 05 49.4	&	1998 Jul 17	&	 900s  (1)	&	300	&	20.81	&	1696-3023	\\
2 & J0055+8518	&	a	&	00 55 37.85	&	85 18 17.3	&	1998 Aug 11	&	1200s (5)	&	150	&	18.25	&	448-5947	\\
3 & J0613+4752	&	a	&	06 12 59.93	&	47 52 43.3	&	1999 Jan 24	&	 900s  (2)	&	400	&	23.30	&	3072-4941	\\
4 & J0622-0118	&	a	&	06 22 13.78	&	-00 18 25.5	&	1999 Jan 24	&	 900s  (2)	&	400	&	23.30	&	2928-4710	\\
5 & J0741+6515	&	a	&	07 41 40.89	&	65 15 23.8	&	1999 Dec 6	&	1500s (1)	&	400	&	22.03	&	2133-3682	\\
  & 	&		&		&		&	1999 Dec 6	&	1200s (2)	&	400	&	22.92	&	2449-4053	\\
6 & 	&	b	&	07 41 45.37	&	65 15 36.2	&	1999 Dec 6	&	1500s (1)	&	400	&	22.03	&	1827-3154	\\
7 & 	&	c	&	07 41 44.67	&	65 15 06.9	&	1999 Dec 6	&	1200s (2)	&	400	&	22.92	&	2864-4854	\\
8 & J0931-0449	&	a	&	09 31 25.97	&	-04 49 45.4	&	1999 May 20	&	1200s (2)	&	300	&	18.04	&	1632-4205	\\
9 & J1039+4145	&	a	&	10 39 41.79	&	41 45 04.4	&	1999 May 20	&	1500s (1)	&	400	&	23.56	&	2393-3924	\\
10 & J1120+2323	&	a	&	11 20 22.36	&	23 22 40.8	&	1999 May 20	&	1186s (1)	&	400	&	23.56	&	2503-4106	\\
11 & 	&	b	&	11 20 23.36	&	23 23 13.0	&	1999 May 20	&	1186s (1)	&	400	&	23.56	&	2797-4588	\\
12 & J1134+0406	&	a	&	11 34 04.71	&	04 06 33.7	&	1999 May 20	&	 900s  (1)	&	300	&	18.51	&	1813-4466	\\
13 & J1237+6219	&	a	&	12 37 06.57	&	62 19 37.1	&	1998 Jul 17	&	 900s  (2)	&	300	&	20.81	&	1195-3362	\\
14 & 	&	c	&	12 37 07.71	&	62 19 24.8	&	1998 Jul 17	&	 900s  (2)	&	300	&	20.81	&	1073-3018	\\
  & 	&		&		&		&	1999 May 20	&	1500s (2)	&	400	&	23.56	&	2257-3702	\\

\enddata																	
\tablenotetext{a}{Units of right ascension are hours, minutes, and seconds, and units of declination are degrees, arcminutes, and arcseconds.}																	
\tablenotetext{b}{Number of exposures combined to get total exposure length is given in parenthesis.}																	
\end{deluxetable}

\clearpage

\begin{deluxetable}{rlrr}							
							
\scriptsize							
\tablenum{2}							
\tablecolumns{4}							
\tablewidth{0pc}							
\tablecaption{Field Obervations}							
\tablehead{							
\cline{1-4} \\							
\colhead{Field} & \colhead{Date} & \colhead{Exposure\tablenotemark{a}} & \colhead{Filter} \\
\colhead{} & \colhead{} & \colhead{(s)} & \colhead{}}							
\startdata							
							
J0040+8505	&	1998 Jul 17	&	1200s (4)	&	I	\\
J0055+8518	&	1998 Jul 17	&	1200s (4)	&	I	\\
J0613+4752	&	1999 Jan 24	&	600s (2)	&	R	\\
J0622-0118	&	\nodata	&	\nodata	&	\nodata	\\
J0741+6515	&	1999 Dec 6	&	3600s (6)	&	R	\\
J0931-0449	&	1999 May 19	&	1000s (4)	&	R	\\
J1039+4145	&	1999 May 19	&	900s (3)	&	R	\\
J1120+2323	&	1999 May 20	&	600s (2)	&	R	\\
J1134+0406	&	1999 May 19	&	450s (2)	&	R	\\
J1237+6219	&	1999 May 19	&	600s (2)	&	R	\\
J1237+6219	&	1998 Jul 17	&	460s (2)	&	I	\\
							
\enddata							
\tablenotetext{a}{Number of exposures combined to get total exposure length are given in parenthesis.}							
\end{deluxetable}							

\clearpage

\begin{deluxetable}{rrllrrrrl}												 			
\scriptsize															
\tablenum{3}															
\tablecolumns{9}															
\tablewidth{0pc}															
\tablecaption{Galaxy Colors}															
\tablehead{															
\cline{1-9} \\															
\colhead{ID Number} & \colhead{Field} & \colhead{Object}& \colhead{z}&\colhead{{\it H}}& \colhead{{\it J-H}} & \colhead{{\it R-H}} & \colhead{{\it I-H}} & \colhead{SLC\tablenotemark{b}} \\
\colhead{} & \colhead{} & \colhead{} & \colhead{} & \colhead{(mag)\tablenotemark{a}} & \colhead{(mag)\tablenotemark{a}} & \colhead{(mag)\tablenotemark{a}} & \colhead{(mag)\tablenotemark{a}} & \colhead{}
}															
\startdata															

1 & J0040+8505	&	a	&	1.63\tablenotemark{c}	&	20.4	&	\nodata	&	\nodata	&	2.6$\pm$0.1	&	2.76	\\
2 & J0055+8518	&	a	&	0.7577	&	19.7	&	\nodata	&	\nodata	&	1.5$\pm$0.1	&	2.76	\\
3 & J0613+4752	&	a	&	1.04\tablenotemark{c}	&	18.1	&	\nodata	&	0.5$\pm$0.3	&	\nodata	&	2.76	\\
4 & J0622-0118	&	a	&	1.1335	&	19.1	&	\nodata	&	\nodata	&	\nodata	&	2.76	\\
5 & J0741+6515	&	a	&	1.4447	&	21.1	&	0.8	&	2.8$\pm$0.1	&	\nodata	&	2.33	\\
6 & 	&	b	&	1.86\tablenotemark{c}	&	22.3	&	0.7	&	1.7$\pm$0.1	&	\nodata	&	2.63	\\
7 & 	&	c	&	1.0612	&	19.6	&	1.0	&	4.2$\pm$0.1	&	\nodata	&	3.44	\\
8 & J0931-0449	&	a	&	0.9794	&	19.0	&	0.7	&	1.8$\pm$0.1	&	\nodata	&	2.58	\\
9 & J1039+4145	&	a	&	1.4905	&	20.2	&	1.4	&	3.5$\pm$0.2	&	\nodata	&	2.73	\\
10 & J1120+2323	&	a	&	1.3825	&	20.5	&	\nodata	&	3.1$\pm$0.1	&	\nodata	&	2.76	\\
11 & 	&	b	&	1.1349	&	21.1	&	\nodata	&	3.2$\pm$0.1	&	\nodata	&	2.76	\\
12 & J1134+0406	&	a	&	0.92\tablenotemark{c}	&	19.9	&	1.0	&	2.8$\pm$0.1	&	\nodata	&	2.76	\\
13 & J1237+6219	&	a	&	1.37\tablenotemark{c}	&	20.5	&	\nodata	&	2.5$\pm$0.4	&	2.6$\pm$0.2	&	2.86	\\
14 & 	&	c	&	1.6377	&	22.3	&	\nodata	&	2.8$\pm$0.1	&	0.7$\pm$0.2	&	2.76	\\


\enddata															
\tablenotetext{a}{Magnitudes based on Vega scale.}															
\tablenotetext{b}{SLC stands for Slit Loss Correction.}	
\tablenotetext{c}{Redshift assuming emission detected in M99 survey is \han.}																													
\end{deluxetable}

\clearpage

\begin{deluxetable}{rrlllllrrrrll}																							
\tabletypesize{\tiny}																							
\tablenum{4}																							
\tablecolumns{13}																							
\tablewidth{0pc}																							
\tablecaption{Emission Line Detections}																							
\tablehead{																							
\cline{1-13} \\																							
\colhead{ID \#} & \colhead{Field} & \colhead{Object} & \colhead{Z$_{\ha}$} & \colhead{Species} & \colhead{\lam$_{rest}$} & \colhead{\lam$_{obs}$} & \colhead{Flux}  &  \colhead{N$_{\sigma}$} & \colhead{Total Error\tablenotemark{a}} & \colhead{EW} & \colhead{Z$_{average}$} & \colhead{Z$_{\sigma}$} \\
\colhead{} & \colhead{} & \colhead{} & \colhead{} & \colhead{} & \colhead{(\AA)} & \colhead{(\AA)} & \colhead{(10$^{-18}$ ergs s$^{-1}$ cm$^{-2}$)} & \colhead{} & \colhead{} & \colhead{(\AA)} & \colhead{} & \colhead{}}																							
\startdata																							

2 & J0055+8518	&	a	&	0.76	&	[O II]	&	2470	&	\nodata	&	$<$32.32	&	\nodata	&	\nodata	&	\nodata	&	0.7579	&	0.0029	\\
 &	&		&		&	Mg II	&	2800	&	\nodata	&	$<$44.13	&	\nodata	&	\nodata	&	\nodata	&		&		\\
 &	&		&		&	O III	&	3133	&	5505.38	&	38.86	&	3.9	&	20.48	&	35.5	&		&		\\
 &	&		&		&	[NeV]	&	3346	&	5881.77	&	18.77	&	2.4	&	11.58	&	11.5	&		&		\\
 &	&		&		&	[O II]	&	3727	&	6561.11	&	98.61	&	11.9	&	46.06	&	70.4	&		&		\\
 &	&		&		&	[Ne III]	&	3869	&	6802.27	&	7.20	&	2.3	&	4.59	&	5.5	&		&		\\
 &	&		&		&	H$\epsilon$	&	3889	&	6846.90	&	7.12	&	2.2	&	4.58	&	5.7	&		&		\\
 &	&		&		&	[Ne III]	&	3967	&	\nodata	&	$<$9.19	&	\nodata	&	\nodata	&	\nodata	&		&		\\
 &	&		&		&	[S II]	&	4072	&	7154.97	&	6.07	&	2.1	&	\nodata	&	5.0	&		&		\\
 &	&		&		&	H$\delta$	&	4102	&	7228.09	&	9.05\tablenotemark{b}	&	2.8	&	5.30	&	8.9	&		&		\\
 &	&		&		&	H$\gamma$	&	4340	&	\nodata	&	$<$8.58\tablenotemark{c}	&	\nodata	&	\nodata	&	\nodata	&		&		\\
 &	&		&		&	[O III]	&	4363	&	\nodata	&	$<$8.94	&	\nodata	&	\nodata	&	\nodata	&		&		\\
 &	&		&		&	He II	&	4686	&	8222.81	&	8.94	&	4.6	&	4.55	&	17.3	&		&		\\
 &	&		&		&	H$\beta$	&	4861	&	8532.68	&	8.11	&	3.0	&	4.62	&	17.1	&		&		\\
 &	&		&		&	[O III]	&	4959	&	8690.77	&	11.95	&	4.3	&	6.16	&	21.2	&		&		\\
 &	&		&		&	[O III]	&	5006	&	8813.18	&	33.26	&	5.0	&	16.65	&	56.2	&		&		\\
4 & J0622-0118	&	a	&	1.14	&	[O II]	&	3727	&	7912.76\tablenotemark{d}	&	79.21\tablenotemark{b}	&	9.2	&	31.03	&	146.1	&	1.1350	&	0.0018	\\
 &	&		&		&	[Ne III]	&	3869	&	8252.90	&	7.89	&	1.5	&	6.12	&	30.1	&		&		\\
 &	&		&		&	H$\epsilon$	&	3889	&	8305.18	&	11.54\tablenotemark{b}	&	2.0	&	7.30	&	20.2	&		&		\\
 &	&		&		&	[Ne III]	&	3967	&	8483.03	&	6.04	&	1.2	&	5.52	&	23.3	&		&		\\
 &	&		&		&	[S II]	&	4072	&	8693.81	&	4.61	&	1.4	&	3.75	&	14.8	&		&		\\
 &	&		&		&	H$\delta$	&	4102	&	8760.22	&	21.14\tablenotemark{b}	&	5.2	&	8.94	&	48.8	&		&		\\
 &	&		&		&	H$\gamma$	&	4340	&	9258.65	&	19.76	&	3.7	&	9.16	&	49.6	&		&		\\
 &	&		&		&	[O III]	&	4363	&	9309.67	&	18.77\tablenotemark{b}	&	2.8	&	8.89	&	58.6	&		&		\\
5 & J0741+6515	&	a	&	1.45	&	[O II]	&	2470	&	6032.56	&	56.90	&	4.7	&	12.16	&	36.7	&	1.4445	&	0.0028	\\
 &	&		&		&	Mg II	&	2800	&	6848.72	&	23.46	&	3.1	&	7.54	&	175.1	&		&		\\
 &	&		&		&	[O II]	&	3727	&	9106.90	&	5.64	&	2.7	&	2.22	&	96.6	&		&		\\
 &	&		&		&	[Ne III]	&	3869	&	\nodata	&	$<$20.60\tablenotemark{c}	&	\nodata	&	\nodata	&	\nodata	&		&		\\
 &	&		&		&	H$\epsilon$	&	3889	&	\nodata	&	$<$20.60\tablenotemark{c}	&	\nodata	&	\nodata	&	\nodata	&		&		\\
7 & J0741+6515	&	c	&	1.06	&	[O II]	&	3727	&	7697.48	&	23.27\tablenotemark{b}	&	2.2	&	10.67	&	98.7	&	1.0612	&	0.0036	\\
 &	&		&		&	[Ne III]	&	3869	&	\nodata	&	$<$38.12\tablenotemark{c}	&	\nodata	&	\nodata	&	\nodata	&		&		\\
 &	&		&		&	H$\epsilon$	&	3889	&	\nodata	&	$<$20.98\tablenotemark{c}	&	\nodata	&	\nodata	&	\nodata	&		&		\\
 &	&		&		&	[S II]	&	4072	&	\nodata	&	$<$28.66	&	\nodata	&	\nodata	&	\nodata	&		&		\\
 &	&		&		&	H$\delta$	&	4102	&	8445.12	&	10.84	&	1.8	&	6.12	&	66.1	&		&		\\
 &	&		&		&	H$\gamma$	&	4340	&	8938.06	&	4.23	&	1.1	&	4.13	&	31.5	&		&		\\
 &	&		&		&	[O III]	&	4363	&	\nodata	&	$<$11.97	&	\nodata	&	\nodata	&	\nodata	&		&		\\
8 & J0931-0449	&	a	&	0.98	&	C III]	&	1909	&	3783.34	&	5848.86	&	54.2	&	107.98	&	112.1	&	0.9775	&	0.0048	\\
 &	&		&		&	[O II]	&	2470	&	\nodata	&	$<$58.82	&	\nodata	&	\nodata	&	\nodata	&		&		\\
 &	&		&		&	Mg II	&	2800	&	5518.58	&	1742.79	&	37.9	&	46.10	&	59.5	&		&		\\
 &	&		&		&	[O II]	&	3727	&	7373.42	&	48.53\tablenotemark{b}	&	4.6	&	10.68	&	9.5	&		&		\\
 &	&		&		&	[Ne III]	&	3869	&	\nodata	&	$<$39.01	&	\nodata	&	\nodata	&	\nodata	&		&		\\
 &	&		&		&	H$\epsilon$	&	3889	&	7724.83\tablenotemark{d}	&	33.49\tablenotemark{b}	&	3.0	&	11.53	&	10.3	&		&		\\
 &	&		&		&	[S II]	&	4072	&	\nodata	&	$<$50.39\tablenotemark{c}	&	\nodata	&	\nodata	&	\nodata	&		&		\\
 &	&		&		&	H$\delta$	&	4102	&	\nodata	&	$<$64.81	&	\nodata	&	\nodata	&	\nodata	&		&		\\
9 & J1039+4145	&	a	&	1.49	&	[O II]	&	2470	&	6127.20	&	9.77	&	3.5	&	2.98	&	17.3	&	1.4812	&	0.0008	\\
 &	&		&		&	Mg II	&	2800	&	\nodata	&	$<$12.18	&	\nodata	&	\nodata	&	\nodata	&		&		\\
 &	&		&		&	[O II]	&	3727	&	9249.51	&	10.67	&	2.5	&	4.32	&	159.8	&		&		\\
 &	&		&		&	[Ne III]	&	3869	&	9668.46\tablenotemark{d}	&	8.60	&	3.9	&	2.42	&	47.3	&		&		\\
 &	&		&		&	H$\epsilon$	&	3889	&	\nodata	&	$<$10.05	&	\nodata	&	\nodata	&	\nodata	&		&		\\
10 & J1120+2323	&	a	&	1.37	&	Mg II	&	2800	&	6682.75	&	95.91	&	14.4	&	35.86	&	55.5	&	1.3829	&	0.0037	\\
 &	&		&		&	[O II]	&	3727	&	8871.89	&	11.81\tablenotemark{b}	&	3.1	&	5.77	&	27.6	&		&		\\
 &	&		&		&	[Ne III]	&	3869	&	\nodata	&	$<$6.40	&	\nodata	&	\nodata	&	\nodata	&		&		\\
 &	&		&		&	H$\epsilon$	&	3889	&	9261.95	&	7.20	&	0.9	&	8.33	&	19.7	&		&		\\
 &	&		&		&	[S II]	&	4072	&	9696.13	&	38.31	&	5.8	&	15.54	&	74.5	&		&		\\
 &	&		&		&	H$\delta$	&	4102	&	\nodata	&	$<$28.12	&	\nodata	&	\nodata	&	\nodata	&		&		\\
11 & J1120+2323	&	b	&	1.13	&	Mg II	&	2800	&	\nodata	&	$<$10.65	&	\nodata	&	\nodata	&	\nodata	&	1.1349	&	0.0036	\\
 &	&		&		&	[O II]	&	3727	&	7971.43	&	11.81\tablenotemark{b}	&	4.6	&	5.04	&	1735.0	&		&		\\
 &	&		&		&	[Ne III]	&	3869	&	8247.52	&	2.48	&	0.9	&	2.89	&	81.4	&		&		\\
 &	&		&		&	H$\epsilon$	&	3889	&	\nodata	&	$<$17.50\tablenotemark{c}	&	\nodata	&	\nodata	&	\nodata	&		&		\\
 &	&		&		&	[S II]	&	4072	&	\nodata	&	$<$14.21	&	\nodata	&	\nodata	&	\nodata	&		&		\\
 &	&		&		&	H$\delta$	&	4102	&	\nodata	&	$<$7.76	&	\nodata	&	\nodata	&	\nodata	&		&		\\
 &	&		&		&	H$\gamma$	&	4340	&	9262.25	&	3.15	&	3.2	&	1.51	&	36.7	&		&		\\
 &	&		&		&	[O III]	&	4363	&	\nodata	&	$<$17.61\tablenotemark{c}	&	\nodata	&	\nodata	&	\nodata	&		&		\\
14 & J1237+6219	&	c	&	1.64	&	Ly$\alpha$	&	1216	&	\nodata	&	$<$1388.53	&	\nodata	&	\nodata	&	\nodata	&	1.6380	&	0.0026	\\
 &	&		&		&	CIV	&	1550	&	4089.51	&	353.91	&	2.9	&	21.68	&	117.8	&		&		\\
 &	&		&		&	C III]	&	1909	&	5040.30	&	46.54	&	2.8	&	2.87	&	201.2	&		&		\\
 &	&		&		&	[O II]	&	2470	&	\nodata	&	$<$9.25	&	\nodata	&	\nodata	&	\nodata	&		&		\\
 &	&		&		&	Mg II	&	2800	&	7373.23	&	6.34\tablenotemark{b}	&	1.1	&	2.21	&	31.3	&		&		\\

\enddata																		
\tablenotetext{a}{Total error includes error in the slit loss correction.  See text for details.}																							
\tablenotetext{b}{Uncertain measurement due to proximity to strong sky emission.}			
\tablenotetext{c}{Feature expected in a region of strong sky emission.}			
\tablenotetext{d}{Observed wavelength greater than expected error, thus only a probably detection and not included in redshift estimate.}																							
\end{deluxetable}										
\clearpage

\begin{deluxetable}{rrllllr}													
\scriptsize													
\tablenum{5}													
\tablecolumns{7}													
\tablewidth{0pc}													
\tablecaption{Upper Limits on Emission Line Fluxes}													
\tablehead{													
\cline{1-7} \\													
\colhead{ID Number} & \colhead{Field} & \colhead{Object}& \colhead{Z$_{\ha}$}&\colhead{Species}& \colhead{\lam$_{rest}$} & \colhead{Flux\tablenotemark{a}} \\
\colhead{} & \colhead{} & \colhead{} & \colhead{} & \colhead{} & \colhead{(\AA)} & \colhead{(10$^{-18}$ ergs s$^{-1}$ cm$^{-2}$)}}													
\startdata													

1	&	J0040+8505	&	a	&	1.63	&	C III]	&	1909	&	$<$27.03	\\
	&		&		&		&	[O II]	&	2470	&	$<$18.42	\\
	&		&		&		&	Mg II	&	2800	&	$<$16.49\tablenotemark{b}	\\
3	&	J0613+4752	&	a	&	1.04	&	[O II]	&	3727	&	$<$117.42	\\
	&		&		&		&	[Ne III]	&	3869	&	$<$106.70\tablenotemark{b}	\\
	&		&		&		&	H$\epsilon$	&	3889	&	$<$106.41	\\
	&		&		&		&	[S II]	&	4072	&	$<$63.64\tablenotemark{b}	\\
	&		&		&		&	H$\delta$	&	4102	&	$<$78.33\tablenotemark{b}	\\
	&		&		&		&	H$\gamma$	&	4340	&	$<$98.92	\\
	&		&		&		&	[O III]	&	4363	&	$<$52.11\tablenotemark{b}	\\
	&		&		&		&	H$\beta$	&	4861	&	$<$158.26	\\
6	&	J0741+6515	&	b	&	1.86	&	C III]	&	1909	&	$<$35.61	\\
	&		&		&		&	[O II]	&	2470	&	$<$16.95	\\
	&		&		&		&	Mg II	&	2800	&	$<$25.92\tablenotemark{b}	\\
12	&	J1134+0406	&	a	&	0.92	&	C III]	&	1909	&	$<$173.07	\\
	&		&		&		&	[O II]	&	2470	&	$<$6.47	\\
	&		&		&		&	Mg II	&	2800	&	$<$8.85	\\
	&		&		&		&	[O II]	&	3727	&	$<$14.10	\\
	&		&		&		&	[Ne III]	&	3869	&	$<$7.57	\\
	&		&		&		&	H$\epsilon$	&	3889	&	$<$7.79\tablenotemark{b}	\\
	&		&		&		&	[S II]	&	4072	&	$<$12.60\tablenotemark{b}	\\
	&		&		&		&	H$\delta$	&	4102	&	$<$12.27\tablenotemark{b}	\\
	&		&		&		&	H$\gamma$	&	4340	&	$<$14.28	\\
	&		&		&		&	[O III]	&	4363	&	$<$17.26\tablenotemark{b}	\\
13	&	J1237+6219	&	a	&	1.37	&	Ly$\alpha$	&	1216	&	$<$2796.58	\\
	&		&		&		&	CIV	&	1550	&	$<$504.03	\\
	&		&		&		&	C III]	&	1909	&	$<$71.60	\\
	&		&		&		&	[O II]	&	2470	&	$<$33.64	\\
	&		&		&		&	Mg II	&	2800	&	$<$32.79	\\

\enddata													
\tablenotetext{a}{All Fluxes are 3$\sigma$ upper limits.}													
\tablenotetext{b}{Feature expected in a region of strong sky emission.}													
\end{deluxetable}													
\clearpage

\begin{deluxetable}{rrllrrrrrrrrrrr}																											
\tabletypesize{\tiny}																											
\rotate																											
\tablenum{6}																											
\tablecolumns{15}												 															
\tablewidth{0pc}																											
\tablecaption{Star Formation Rates}																											
\tablehead{																											
\cline{1-15} \\																											
\colhead{ID \#} & \colhead{Field} & \colhead{Object} & \colhead{z} & \colhead{SFR$_{\ha}$\tablenotemark{a}} & \colhead{L$_{[OII]}$} & \colhead{M$_{B}$} & \colhead{log([O II]/\han$_{pred}$)} & \colhead{log([O II]/\han$_{obs}$)} & \colhead{SFR$_{[OII]}$\tablenotemark{a}} & \colhead{SFR$_{[OII]}$\tablenotemark{b}} & \colhead{SFR$_{[OII]}$\tablenotemark{c}} &  \colhead{SFR$_{[OII]}$\tablenotemark{d}} & \colhead{E(B-V)$_{[O II]}$\tablenotemark{e}} & \colhead{E(B-V)$_{Bal}$\tablenotemark{f}} \\	
\colhead{} & \colhead{} & \colhead{} & \colhead{} & \colhead{(M$_{\odot}$yr$^{-1}$)} & \colhead{(10$^{40}$ergs s$^{-1}$)} & \colhead{} & \colhead{} & \colhead{} & \colhead{(M$_{\odot}$yr$^{-1}$)} & \colhead{(M$_{\odot}$yr$^{-1}$)} & \colhead{(M$_{\odot}$yr$^{-1}$)} & \colhead{(M$_{\odot}$yr$^{-1}$)} & \colhead{} & \colhead{} \\						
\colhead{(1)} & \colhead{(2)} & \colhead{(3)} & \colhead{(4)} & \colhead{(5)} & \colhead{(6)} & \colhead{(7)} & \colhead{(8)} & \colhead{(9)} & \colhead{(10)} & \colhead{(11)} & \colhead{(12)} & \colhead{(13)} & \colhead{(14)} & \colhead{(15)}}																											
\startdata																											

1 & J0040+8505	&	a	&	1.63\tablenotemark{g}	&	16	&	\nodata	&	-21.89	&	-0.36	&	\nodata	&	\nodata	&	\nodata	&	\nodata	&	\nodata	&	\nodata	&	\nodata	\\
2 & J0055+8518	&	a	&	0.7579	&	10	&	31.78	&	-21.93	&	-0.37	&	 -0.56 $^{+0.18}_{-0.30}$	&	6.3	&	15.8	&	16.5	&	40.9	&	0.21 $^{+ 0.20}_{- 0.35}$	&	1.5	\\
3 & J0613+4752	&	a	&	1.04\tablenotemark{g}	&	5	&	$<$75.91	&	-24.27	&	-0.57	&	$<$0.13	&	$<$15.0	&	$<$37.7	&	$<$63.0	&	$<$2.5	&	$>$-0.77	&	$>$-1.4	\\
4 & J0622-0118	&	a	&	1.1350	&	13	&	62.14	&	\nodata	&	\nodata	&	 -0.39 $^{+0.15}_{-0.23}$	&	12.3	&	30.9	&	\nodata	&	\nodata	&	\nodata	&	0.0	\\
5 & J0741+6515	&	a	&	1.4445	&	8	&	7.56	&	-20.59	&	-0.25	&	 -1.07 $^{+0.13}_{-0.19}$	&	1.5	&	3.8	&	3.0	&	152.3	&	0.91 $^{+ 0.17}_{- 0.25}$	&	$>$-1.8	\\
6 & 	&	b	&	1.86\tablenotemark{g}	&	25	&	\nodata	&	-20.45	&	-0.24	&	\nodata	&	\nodata	&	\nodata	&	\nodata	&	\nodata	&	\nodata	&	\nodata	\\
7 & 	&	c	&	1.0612	&	7	&	15.74	&	-21.87	&	-0.36	&	 -0.71 $^{+0.17}_{-0.27}$	&	3.1	&	7.8	&	8.1	&	43.4	&	0.39 $^{+ 0.18}_{- 0.30}$	&	0.5	\\
8 & J0931-0449\tablenotemark{i}	&	a	&	0.9775	&	81	&	$<$27.36	&	-22.33	&	-0.40	&	 -1.52 $^{+0.12}_{-0.17}$	&	$<$5.4	&	$<$13.6	&	$<$15.4	&	$<$3155.9	&	1.24 $^{+ 0.10}_{- 0.13}$	&	0.5	\\
9 & J1039+4145	&	a	&	1.4812	&	15	&	15.14	&	-21.11	&	-0.29	&	 -1.05 $^{+0.16}_{-0.26}$	&	3.0	&	7.5	&	6.7	&	245.8	&	0.84 $^{+ 0.17}_{- 0.26}$	&	$>$-0.2	\\
10 & J1120+2323	&	a	&	1.3829	&	14	&	14.39	&	-20.89	&	-0.28	&	 -1.03 $^{+0.19}_{-0.34}$	&	2.8	&	7.2	&	6.1	&	221.5	&	0.84 $^{+ 0.20}_{- 0.35}$	&	0.3	\\
11 & 	&	b	&	1.1349	&	7	&	9.27	&	-19.58	&	-0.16	&	 -0.90 $^{+0.17}_{-0.28}$	&	1.8	&	4.6	&	3.0	&	100.7	&	0.82 $^{+ 0.18}_{- 0.29}$	&	1.0	\\
12 & J1134+0406	&	a	&	0.92\tablenotemark{g}	&	16	&	$<$6.96	&	-20.43	&	-0.24	&	$<$-1.41	&	$<$1.4	&	$<$3.5	&	$<$2.7	&	$<$686.2	&	$>$1.30	&	$>$0.6	\\
13 & J1237+6219	&	a	&	1.37\tablenotemark{g}	&	13	&	\nodata	&	-20.94	&	-0.28	&	\nodata	&	\nodata	&	\nodata	&	\nodata	&	\nodata	&	\nodata	&	\nodata	\\
14 & 	&	c	&	1.6380	&	4	&	\nodata	&	-19.71	&	-0.17	&	\nodata	&	\nodata	&	\nodata	&	\nodata	&	\nodata	&	\nodata	&	\nodata	\\

\enddata																											

\tablenotetext{a}{SFR with [N II] correction applied as described in the text and no extinction correction.}
\tablenotetext{b}{SFR with [N II] correction applied as described in the text and assuming E(\han) = 1 magnitude.}
\tablenotetext{c}{SFR with [N II] correction applied as described in the text, assuming [O II]/\ha value predicted by Jansen et al. (2001), see column 8, and E(\han) = 1 magnitude.}
\tablenotetext{d}{SFR with [N II] correction applied as described in the text, assuming [O II]/\ha value predicted by Jansen et al. (2001), see column [8], and assuming reddening given in column [14] in addition to E(\han) = 1 mag.  See text for how the addtional extinction was derived.}
\tablenotetext{e}{Reddening based on predicted [O II]/\ha value given by Jansen et al. (2001) and observed value.  See text for details.}
\tablenotetext{f}{Upper limit on reddening measured from Balmer lines.}
\tablenotetext{g}{Redshift assuming emission detected in M99 survey is \han.}
\tablenotetext{h}{Seyfert 1 galaxy, thus L([O II]) and SFRs are upper limits.}
\end{deluxetable}

\clearpage
\begin{figure}
\figurenum{1}
\centering
\plotone{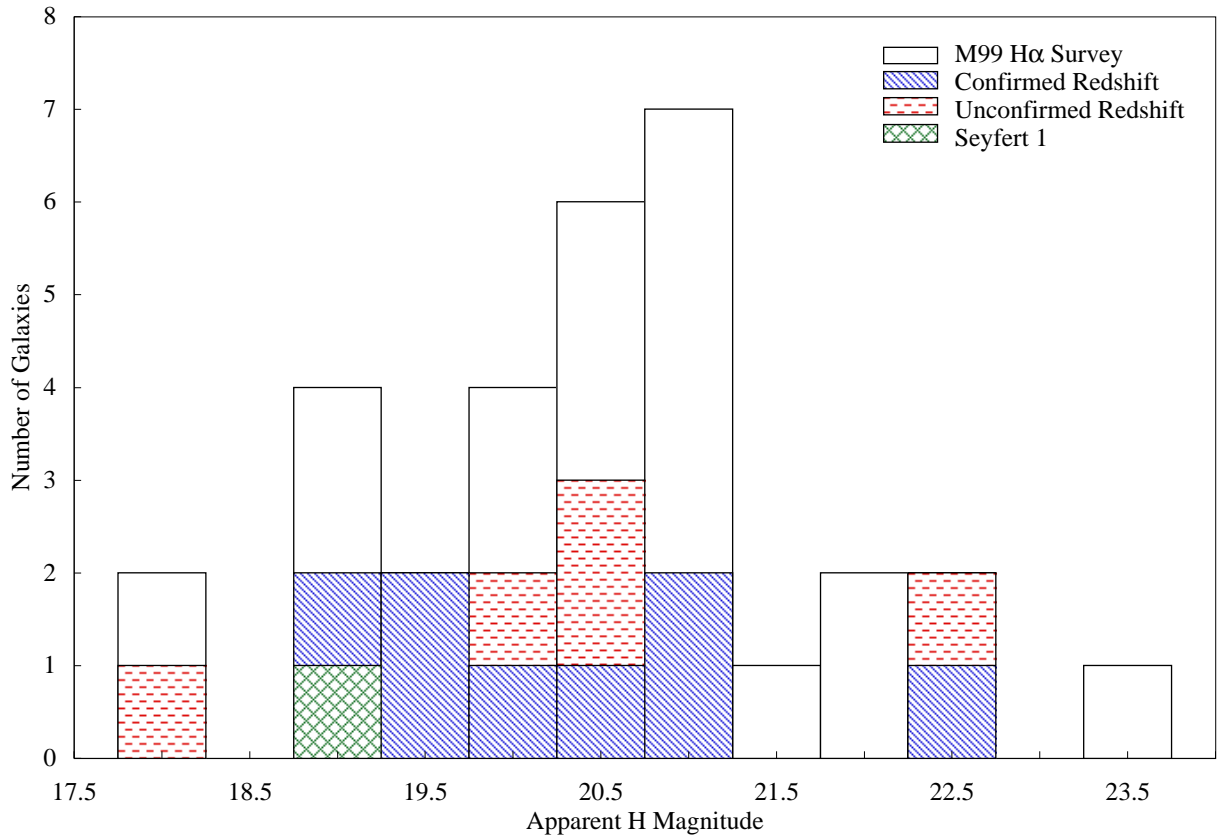}
\caption[]{Distribution of the apparent magnitudes of emission-line galaxies.  Shown in white are those galaxies in the M99 \ha survey for which optical spectra have not been obtained in this study.  The squares filled with blue diagonal lines represent those emission-line galaxies for which we have obtained redshift confirmations, and squares with red horizontal dashes are galaxies for which redshifts were not determinable (see text for explanations).  The single square with green crossed lines is the galaxy J0931-0449a, classified as a Seyfert 1.  The magnitudes are in the Johnson system.}
\end{figure}

\begin{figure}
\figurenum{2}
\centering
\plotone{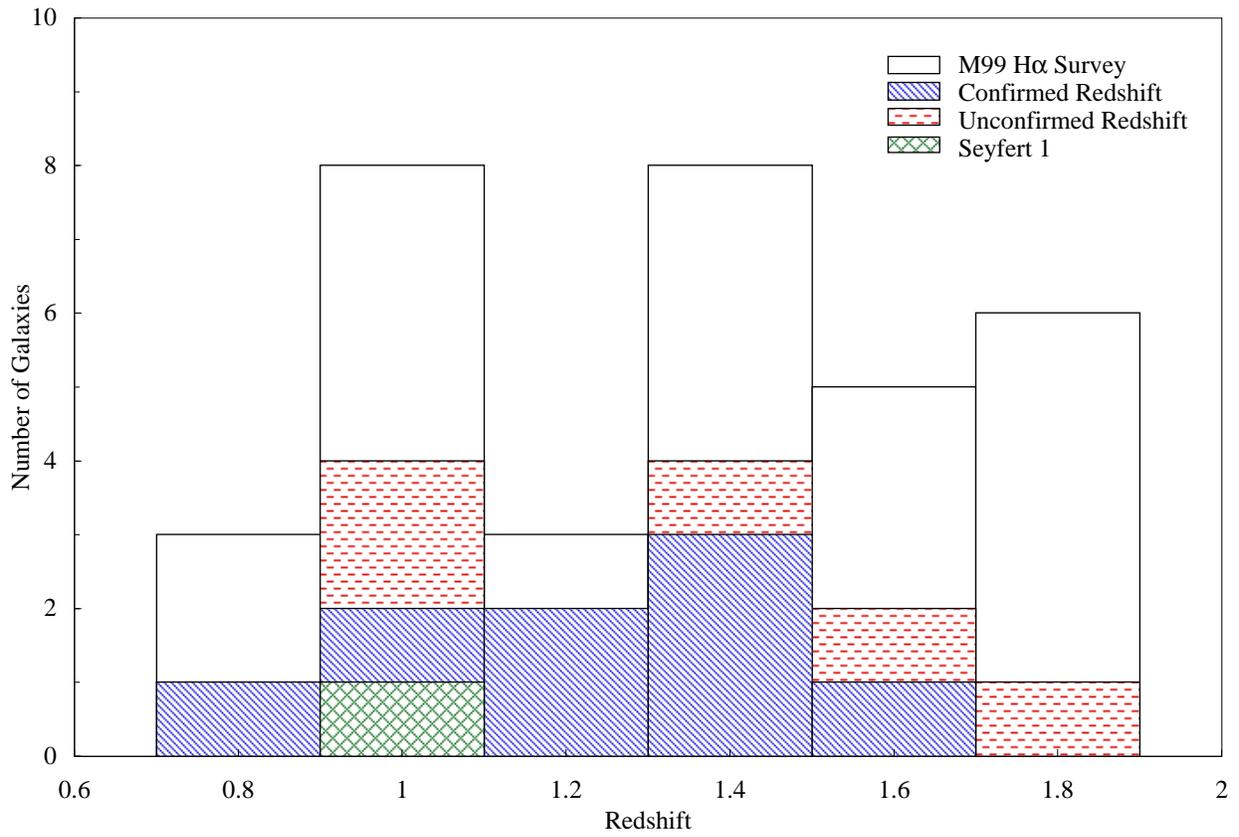}
\caption[]{Distribution of redshifts of the emission-line galaxies.  Regions shaded as in Fig. 1.}
\end{figure}

\begin{figure}
\figurenum{3}
\centering
\plotone{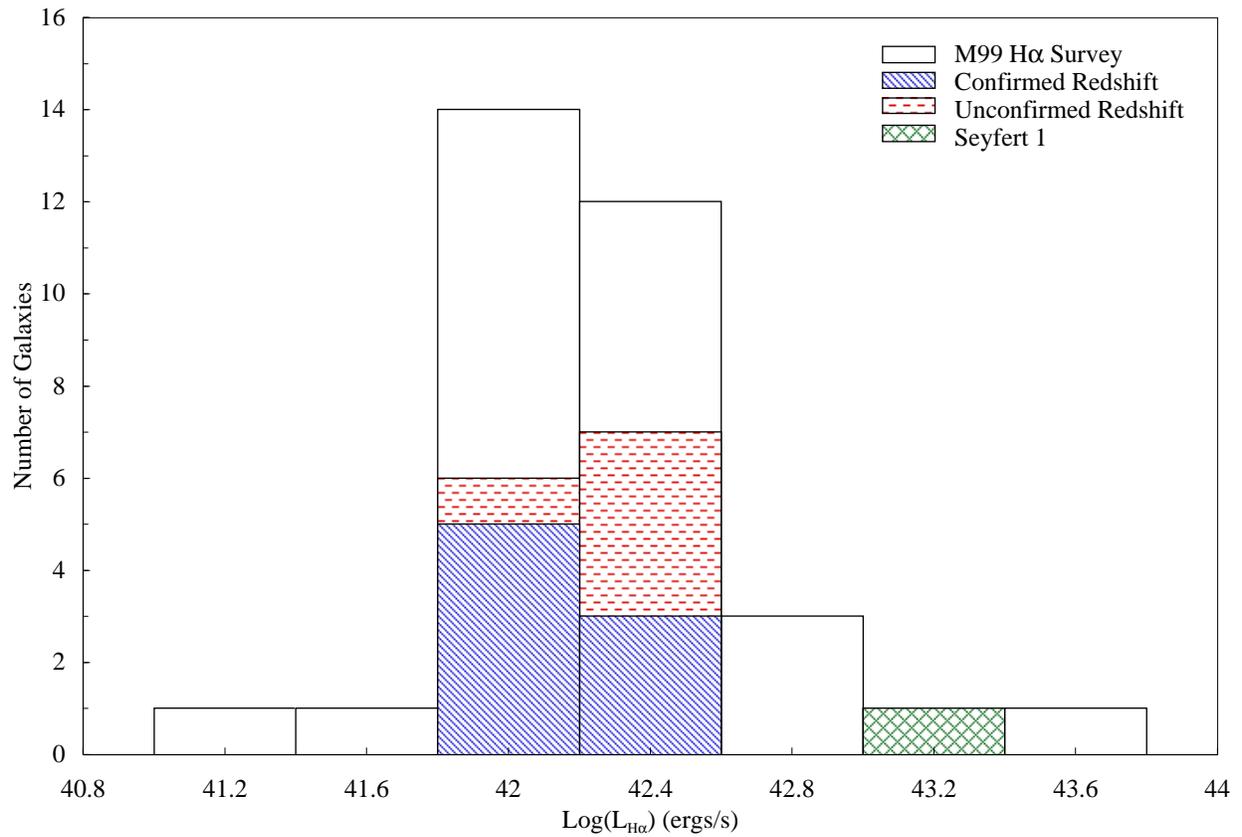}
\caption[]{Distribution of the \ha line luminosities of the emission-line galaxies.  Regions shaded as in Fig. 1.}

\end{figure}

\clearpage
\begin{figure}
\figurenum{4a}
\plotone{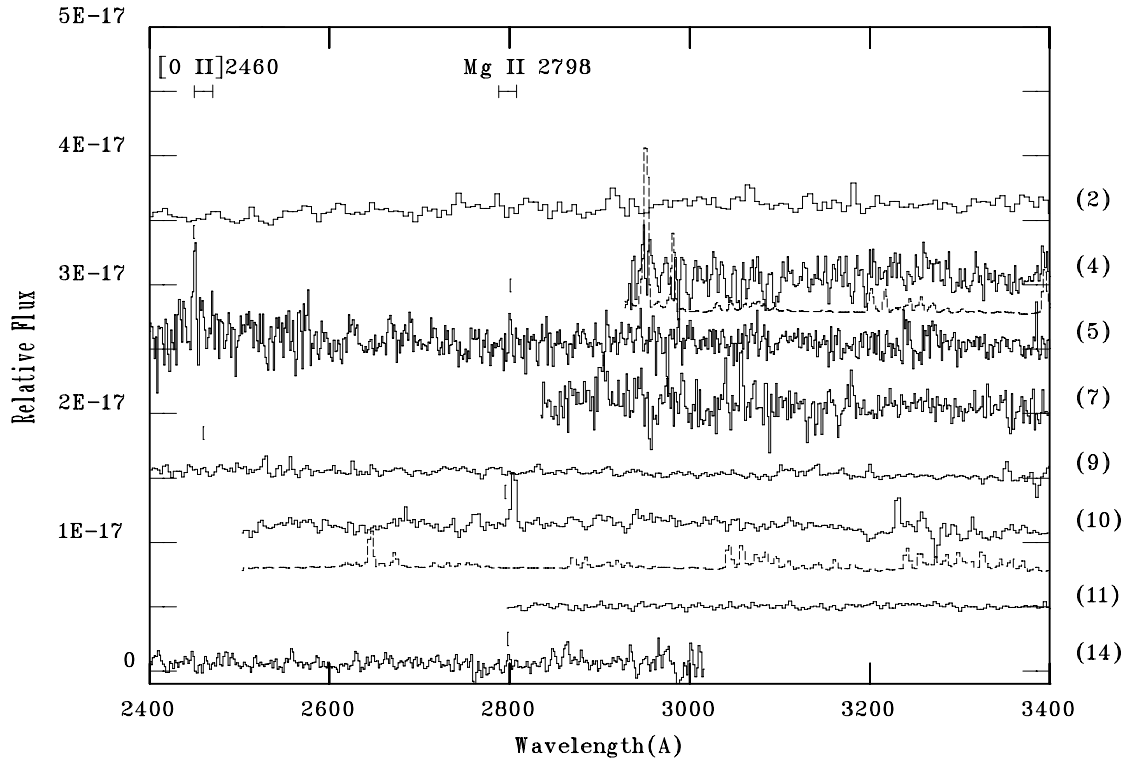}
\caption[]{(a) Confirmed galaxy spectra from 2400 to 3400 \AA, containing the [O II] \lam2470 and Mg II \lam2798 emission features.  (b) Confirmed galaxy spectra from 3400 to 4400 \AA, containing the [O II] \lam3727, [Ne III] \lam3869, H$\epsilon$, H$\delta$, and H$\gamma$ emission features.  Each galaxy spectrum is represented by a solid line histogram spectrum.  The y-axis is in units of ergs cm$^{-2}$ s$^{-1}$ \AA$^{-1}$ and each galaxy spectrum scaled by 2 as well as separated by 5 x 10$^{-18}$ ergs cm$^{-2}$ s$^{-1}$ \AA$^{-1}$.  Thus, each tick mark on the y-axis represents the zero point for the corresponding galaxy spectrum.  Each spectrum has been redshifted to the rest frame.  The redshifts used are those in Table 4 and are based on all detected lines, unless otherwise noted.  Thus due to our error in wavelength, slight shifts in the position of a given feature in different objects are noticeable.  The two dashed spectra are example sky spectra that were subtracted from the galaxy spectra immediately above.  The top dashed spectrum is the sky spectrum used for J0622-0018a and the bottom is the sky spectrum used for J1120+2323a.  The numbers to the right identifies the spectrum to the left, with the number identifications as listed in Table 1.  The emission features present in each spectral range, and the window each was measure within, are labeled at the top of each plot.  Vertical lines indicate a detection of the corresponding emission feature in the spectrum immediately below.}
\end{figure}
\clearpage

\begin{figure}
\figurenum{4b}
\plotone{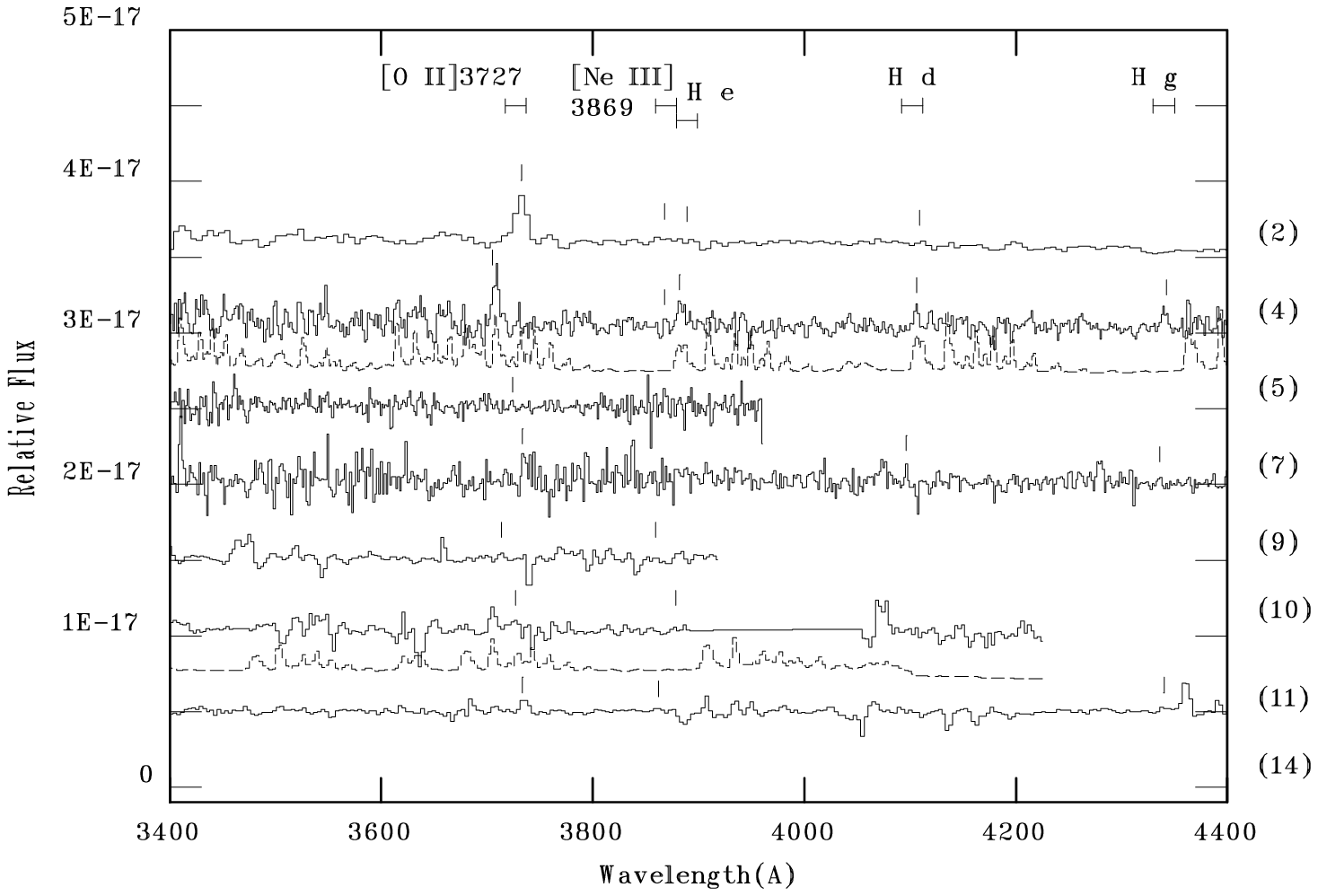}
\caption[]{}
\end{figure}
\clearpage

\begin{figure}
\figurenum{5a}
\centering
\plotone{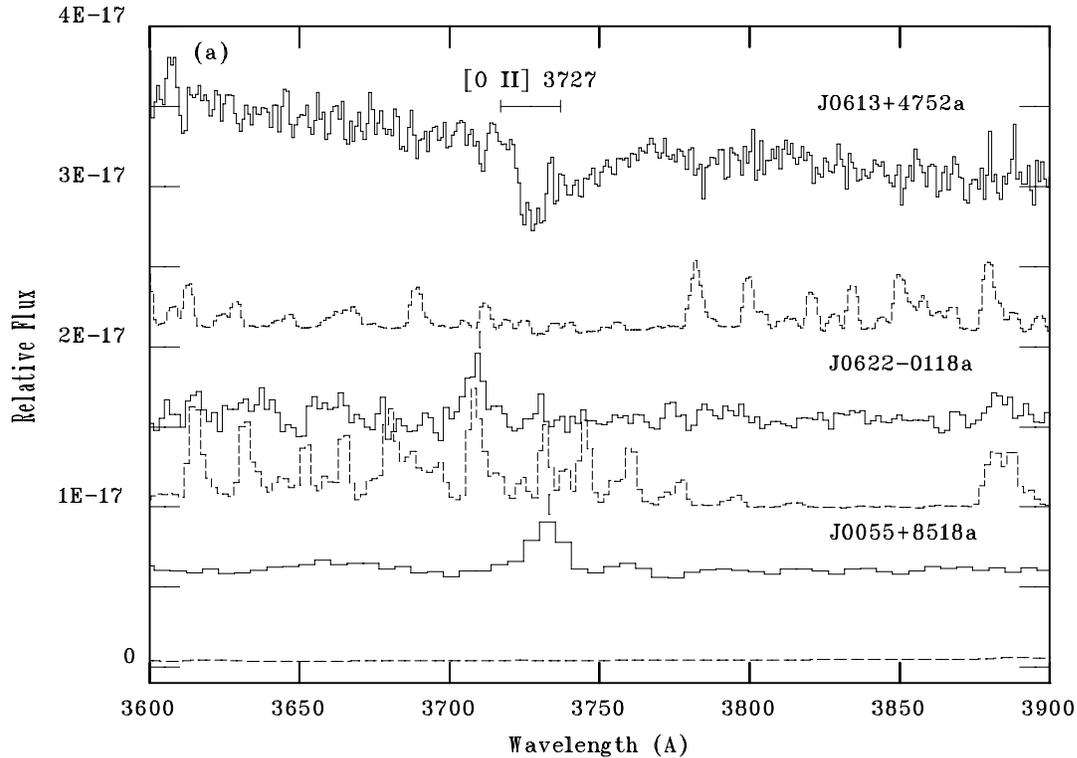}
\caption[]{Spectra of each galaxy for which [O II] \lam 3727 was observed.  For those galaxies with no [O II] detection, the region expected to contain [O II] based on the redshift given by M99, is shown.  J0931-0449a is not included in this figure, since it is a Seyfert 1.  Galaxy spectra are represented by a solid line and the sky spectrum used to subtract sky emission features is the dashed line spectrum immediately below the galaxy spectrum.  The y-axis is in units of ergs cm$^{-2}$ s$^{-1}$ \AA$^{-1}$.  Each spectrum has been redshifted to the rest frame.  The redshifts used are those in Table 4 and are based on all detected lines, unless otherwise noted, thus due to our error in wavelength, slight shifts in the position of the [O II] features in different objects are noticeable.  Each galaxy spectrum is scaled by 2, with the exception of J0613+4752a, which is scaled by 0.5 due to the strong stellar contamination in the spectrum resulting in a significant slope in the continuum.  The sky spectra are scaled by 0.1 and all spectra are offset by 5 x 10$^{-18}$ ergs cm$^{-2}$ s$^{-1}$ \AA$^{-1}$.  Detections of [O II] are marked with a small tick above the feature.}
\end{figure}

\begin{figure}
\figurenum{5b}
\centering
\plotone{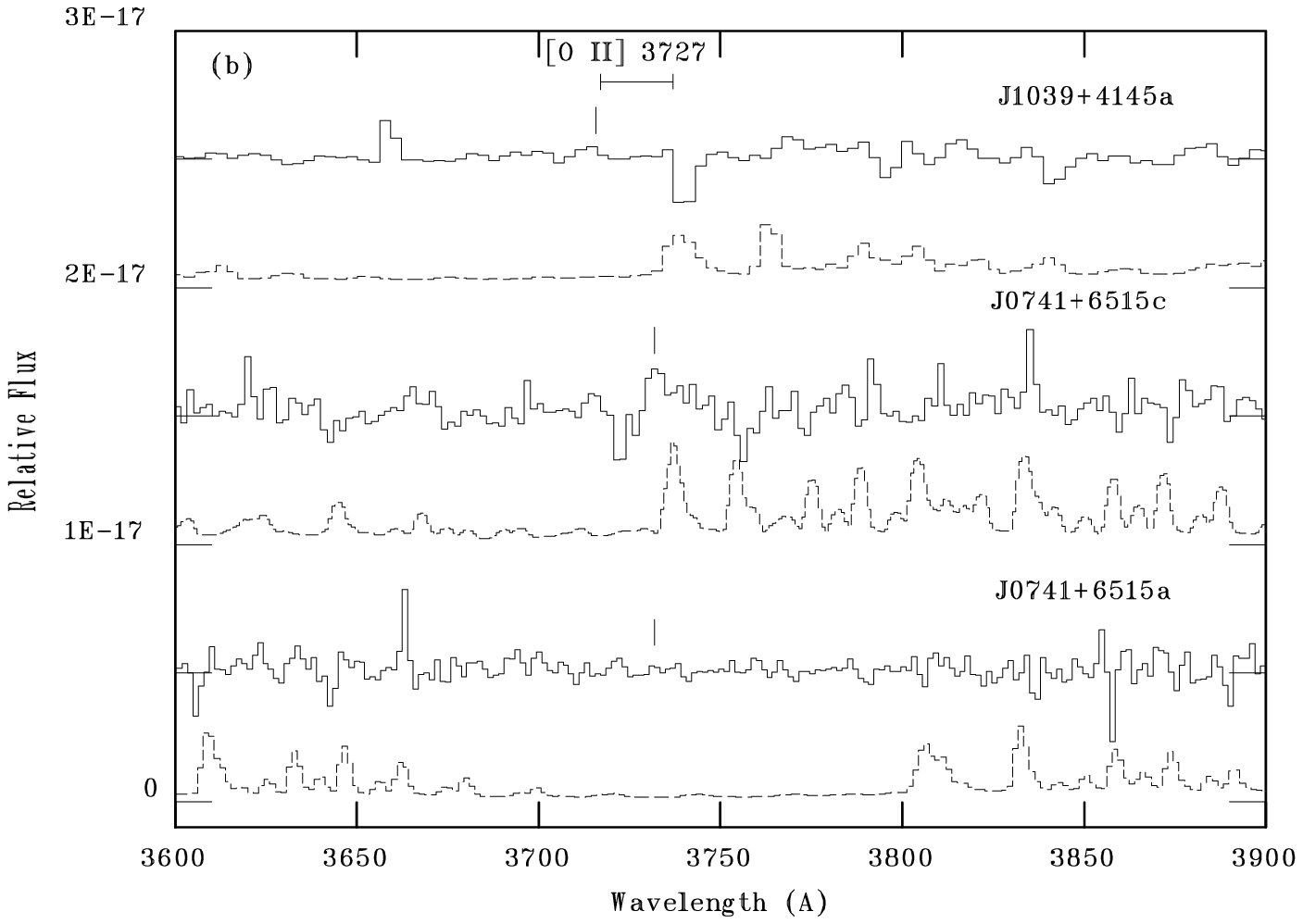}
\caption[]{}
\end{figure}

\begin{figure}
\figurenum{5c}
\centering
\plotone{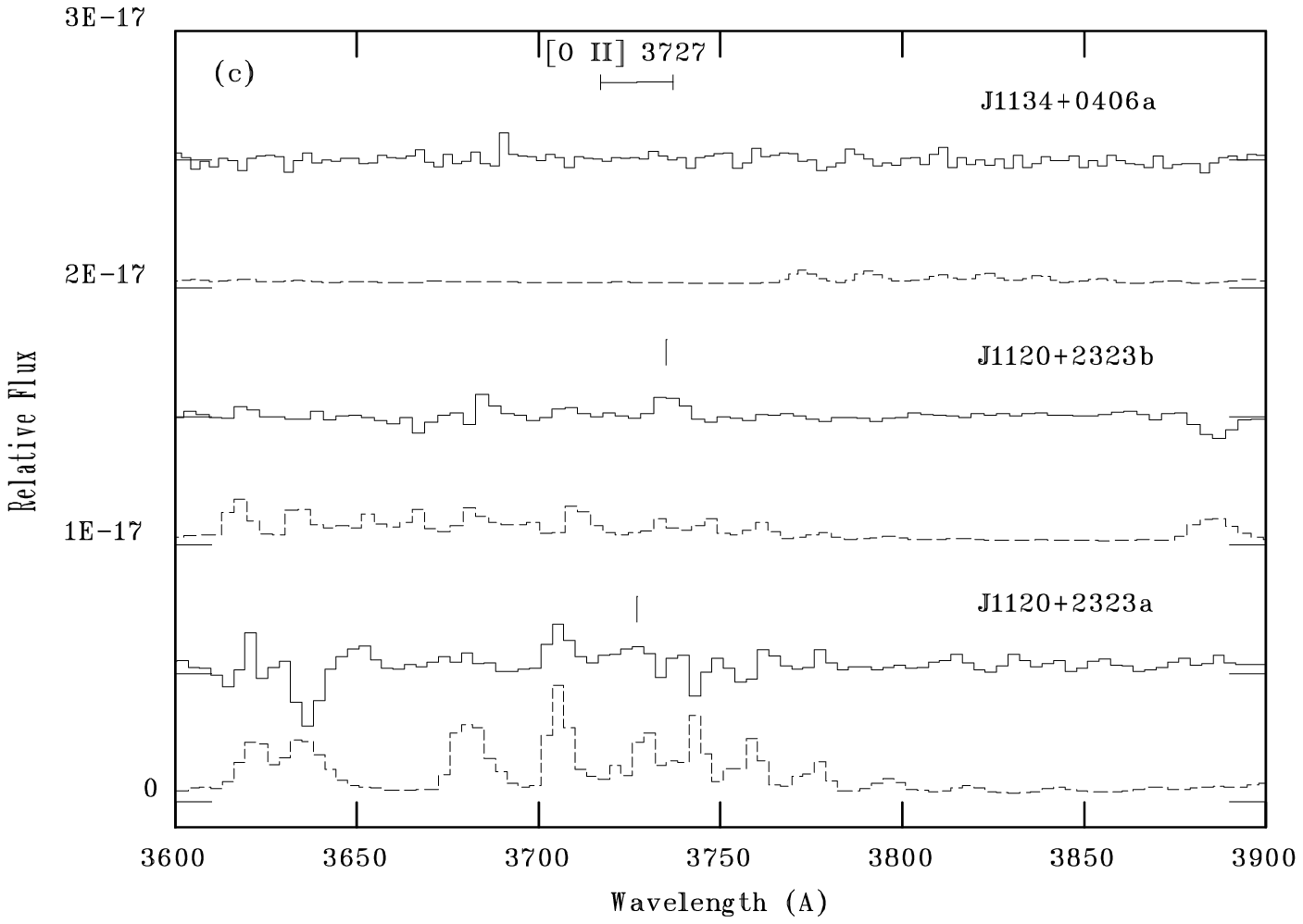}
\caption[]{}
\end{figure}

\begin{figure}
\figurenum{6}
\centering
\plotone{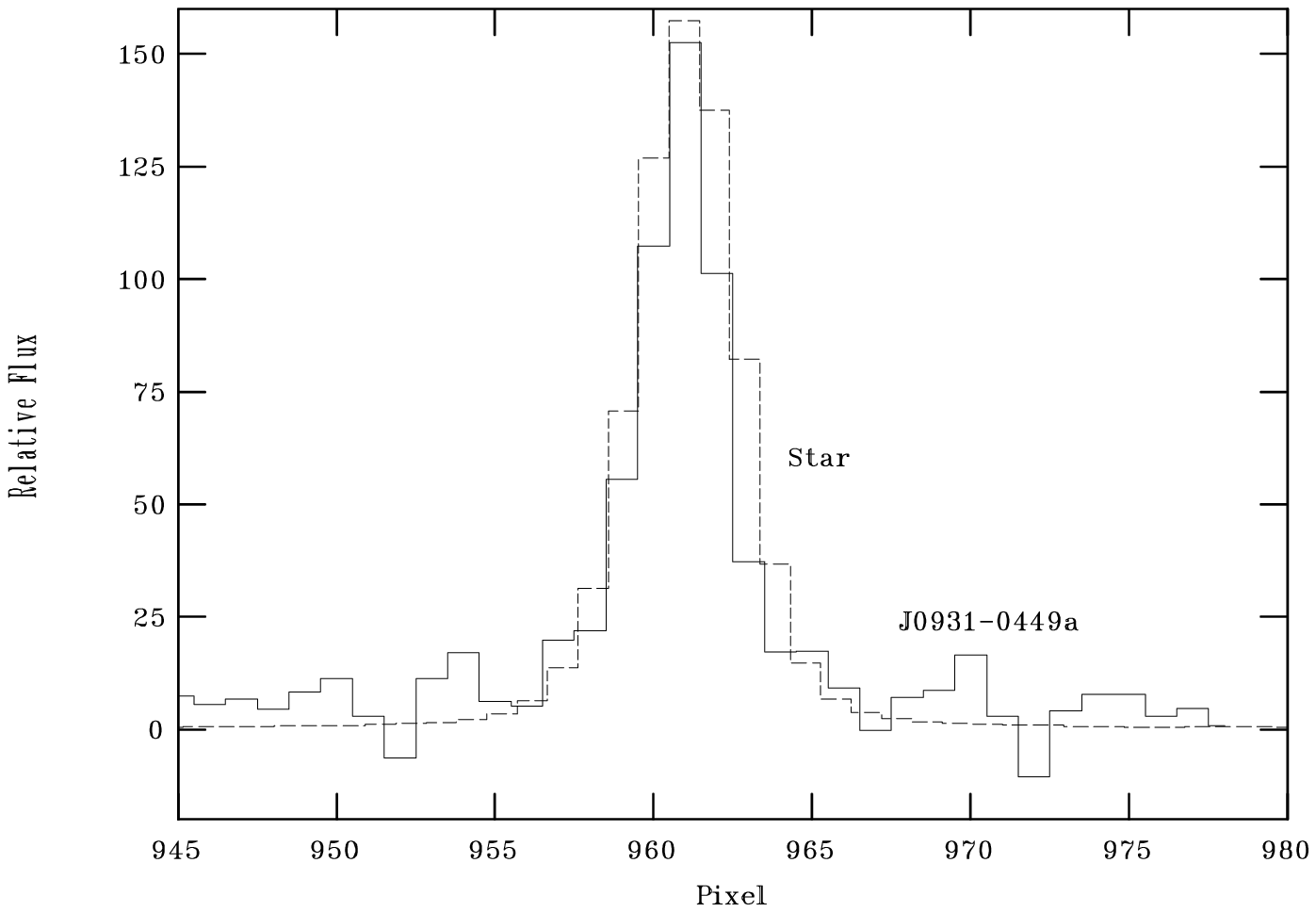}
\caption[]{A spatial cut of the broad emission line C III] in the Seyfert 1 J0931-0449a.  The solid line is the galaxy, and the dashed line is a star, both with a pixel scale of 0$''$.15 pixel$^{-1}$.  The profile of the broad emission is consistent with the emitting region being a point source.}
\end{figure}

\begin{figure}
\figurenum{7}
\centering
\plotone{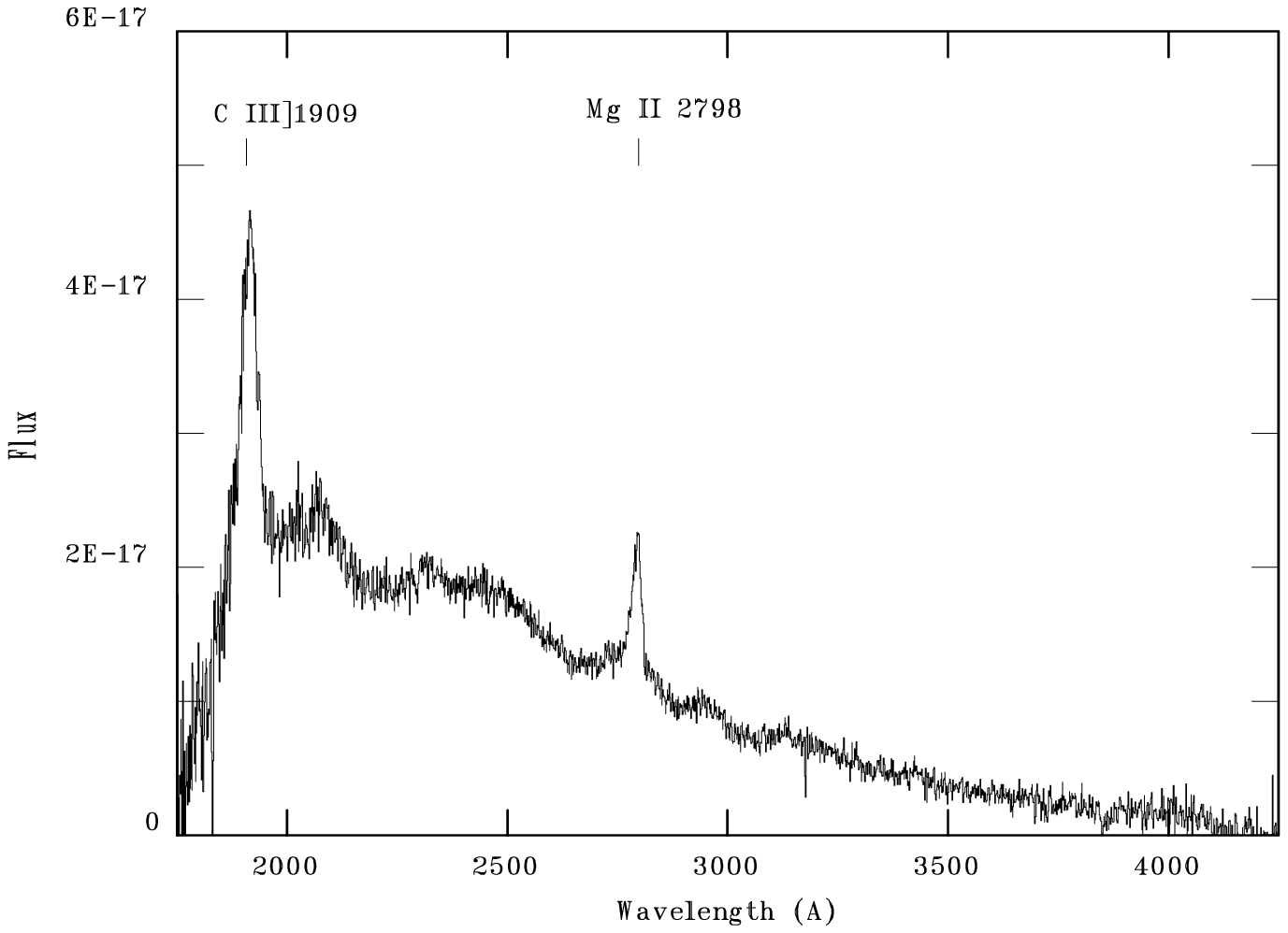}
\caption[]{Complete spectrum of the Seyfert 1 galaxy J0931-0449a. The y-axis is the flux scale in units of cm$^{-2}$ s$^{-1}$ \AA$^{-1}$.  The two strong broad emission lines are C III] \lam1909 and Mg II \lam2798 at z = 0.9775.}
\end{figure}

\begin{figure}
\figurenum{8}
\centering
\plotone{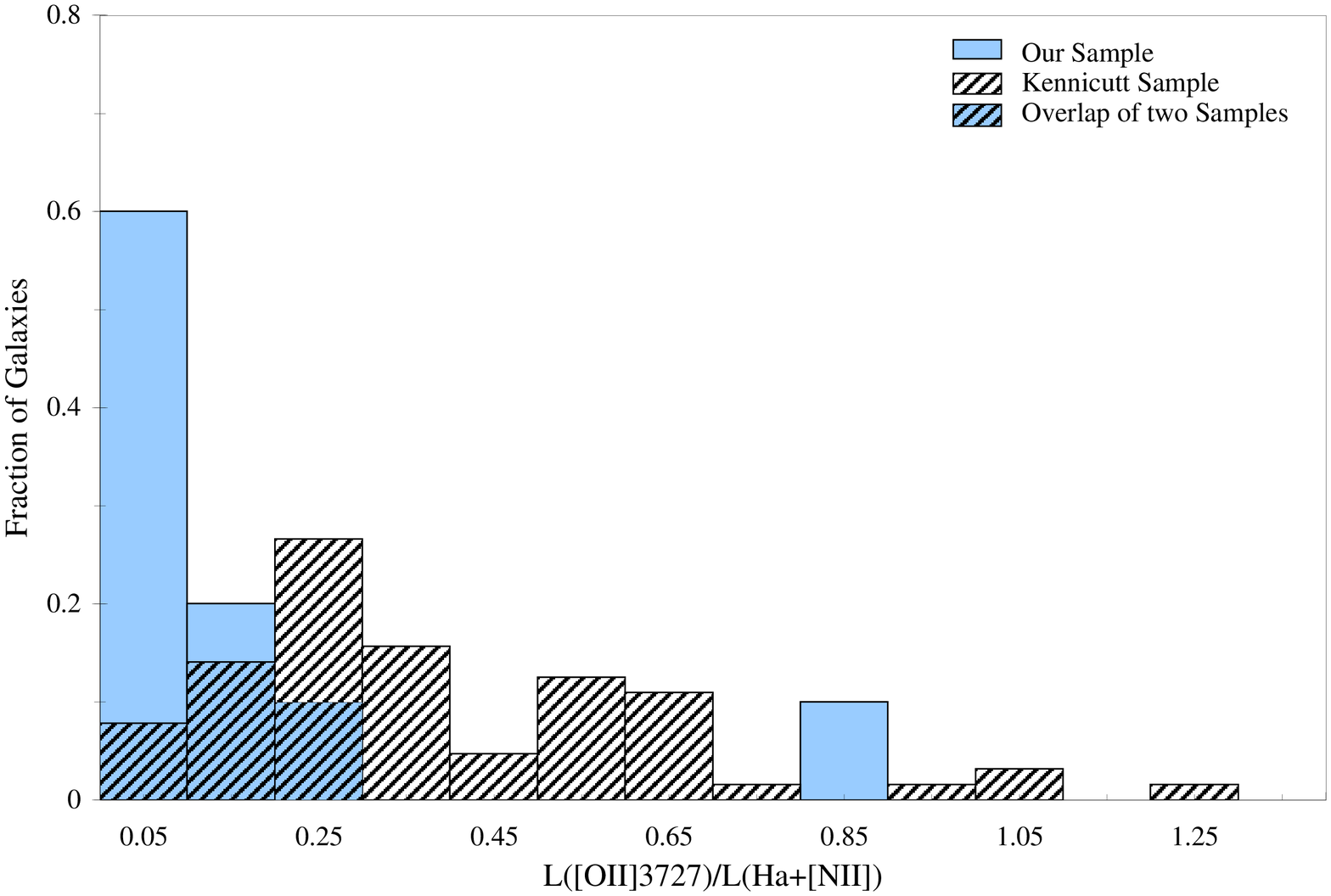}
\caption[]{Distribution of [O II]/\ha in the Kennicutt (1992) sample (black diagonal lines) and our sample (blue), with regions where the samples overlap represented by blue with diagonal lines.  Both samples have been normalized such that the y-axis represents the fraction of galaxies in the total sample.}

\end{figure}

\begin{figure}
\figurenum{9}
\centering
\plotone{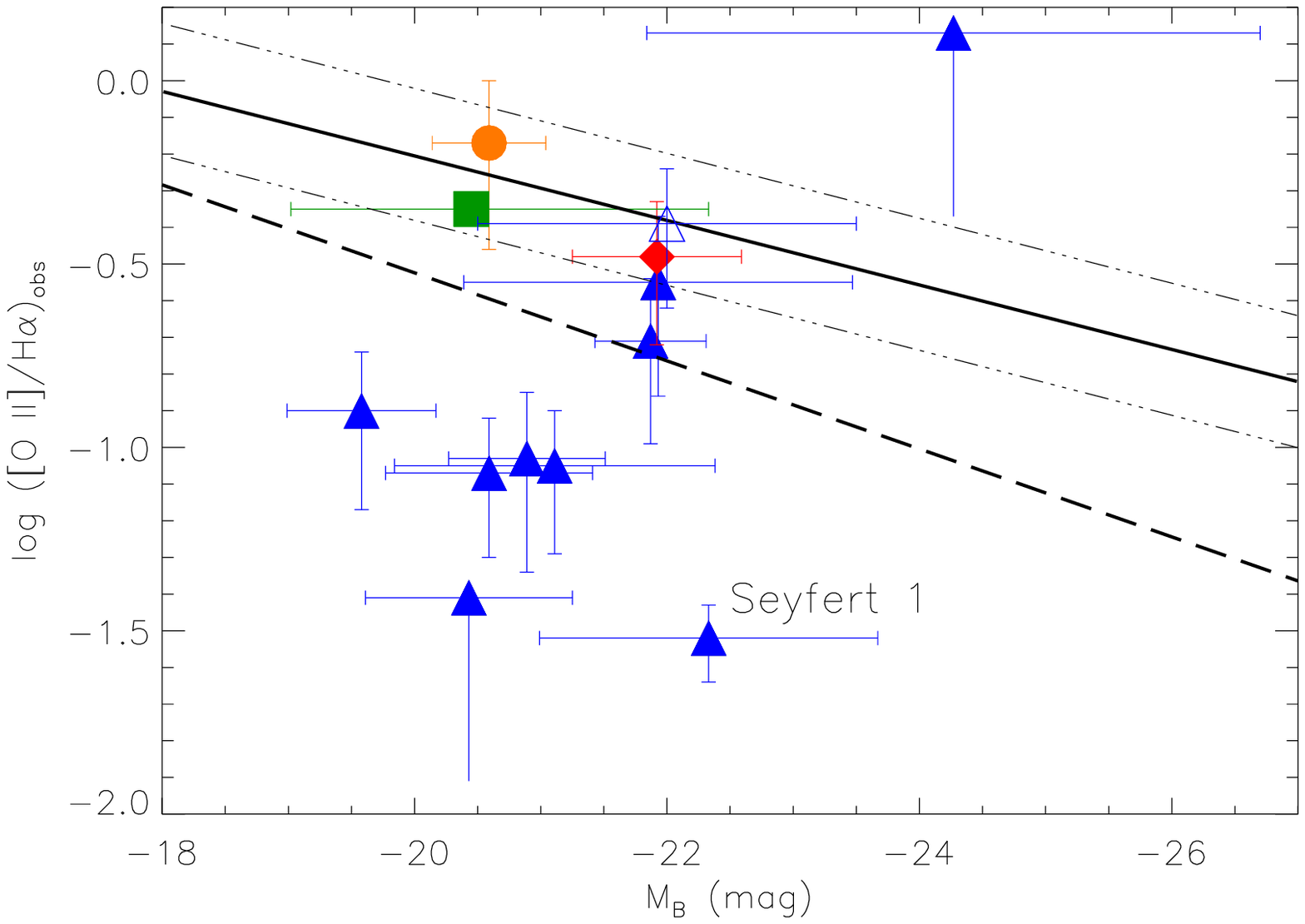}
\caption[]{Logarithm of the observed [O II]/\ha vs. total absolute {\it B} magnitude.  The heavy solid line shows the fit found by Jansen et al. (2001) to a sample of nearby galaxies.  The parallel light dashed lines represent the total scatter in their data.  The dashed heavy line is the fit to local galaxies found by Arag\'{o}n-Salamanca et al. (2002).  Our ten galaxies with [O II] measurements are included as blue triangles, with the Seyfert 1 labeled as such.   J0622-0018a is included in the plot as an open triangle, even though its M$_{B}$ measurement is uncertain (see text for explanation), which is indicated by the large error bars.  Most of our objects are inconsistent with the local relationship found by Jansen et al. (2001).  The green square is the H98 sample, assuming their data has [O II]/\ha = 0.45.  The red diamond and orange circle represent the results of the Pettini et al. (2001) and Glazebrook et al. (1999) studies, respectively.}
\end{figure}


\begin{figure}
\figurenum{10}
\centering
\plotone{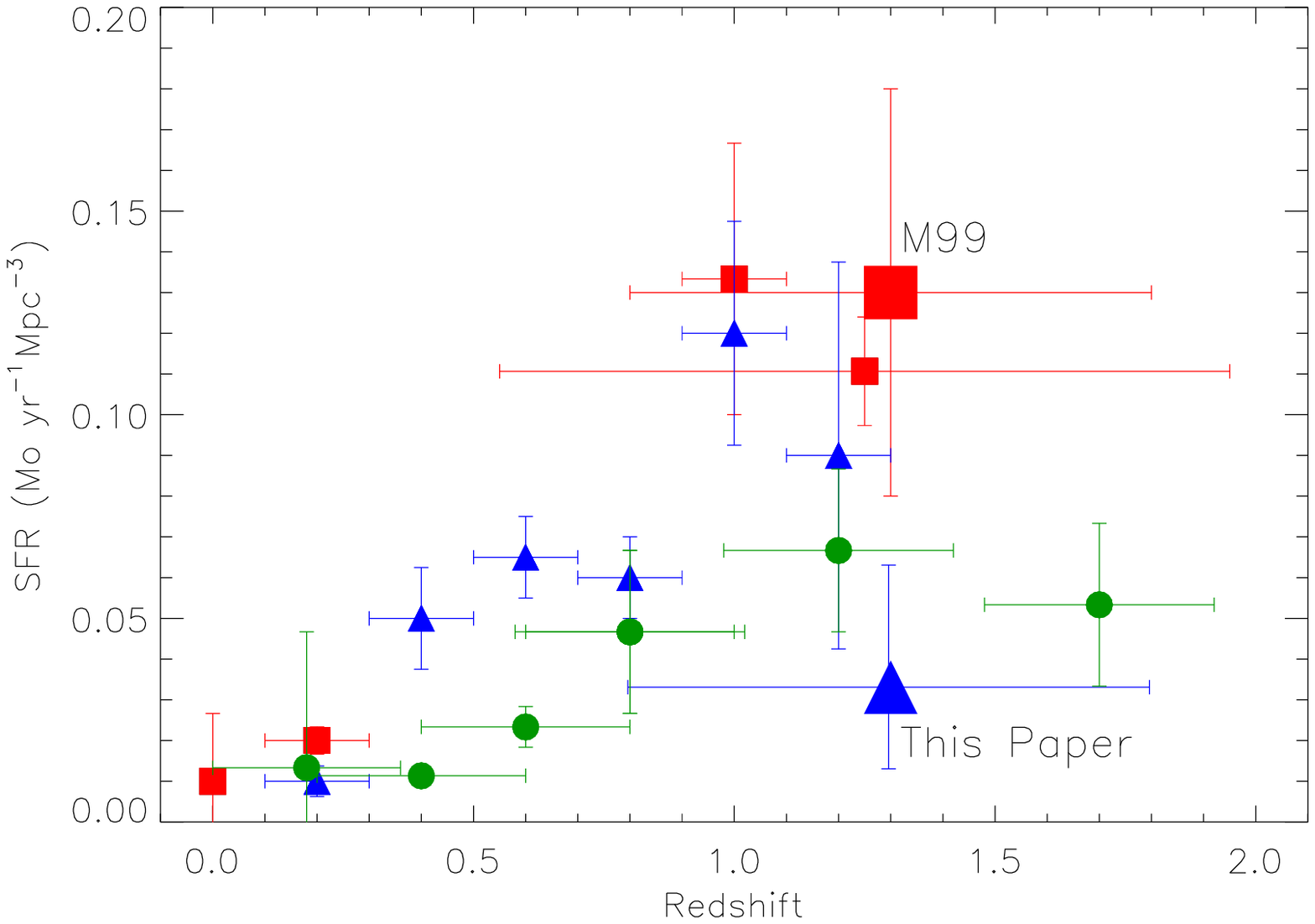}
\caption[]{Global volume-averaged SFR as a function of redshift, uncorrected for dust extinction.  This plot is a compilation of SFR densities derived from emission-line and UV continuum measurements taken from the literature.  Red squares are studies based on \ha (Gallego et al. 1995; Tresse \& Maddox 1998; Glazebrook et al. 1999; Yan et al. 1999; Hopkins, Connolly \& Szalay 2000), with the results found by Yan et al., which are based on the M99 sample, represented by the larger square.  Measurements from studies based on [O II] are shown with blue triangles (H98; this study), with the larger triangle representing the result of this study.  No correction has been made for the additional reddening present in our sample.  Green circles are based on UV continuum measurements (Lilly et al. 1996; Connolly et al. 1997; Treyer et al. 1998).}
\end{figure}

\begin{figure}
\figurenum{11a}
\centering
\plotone{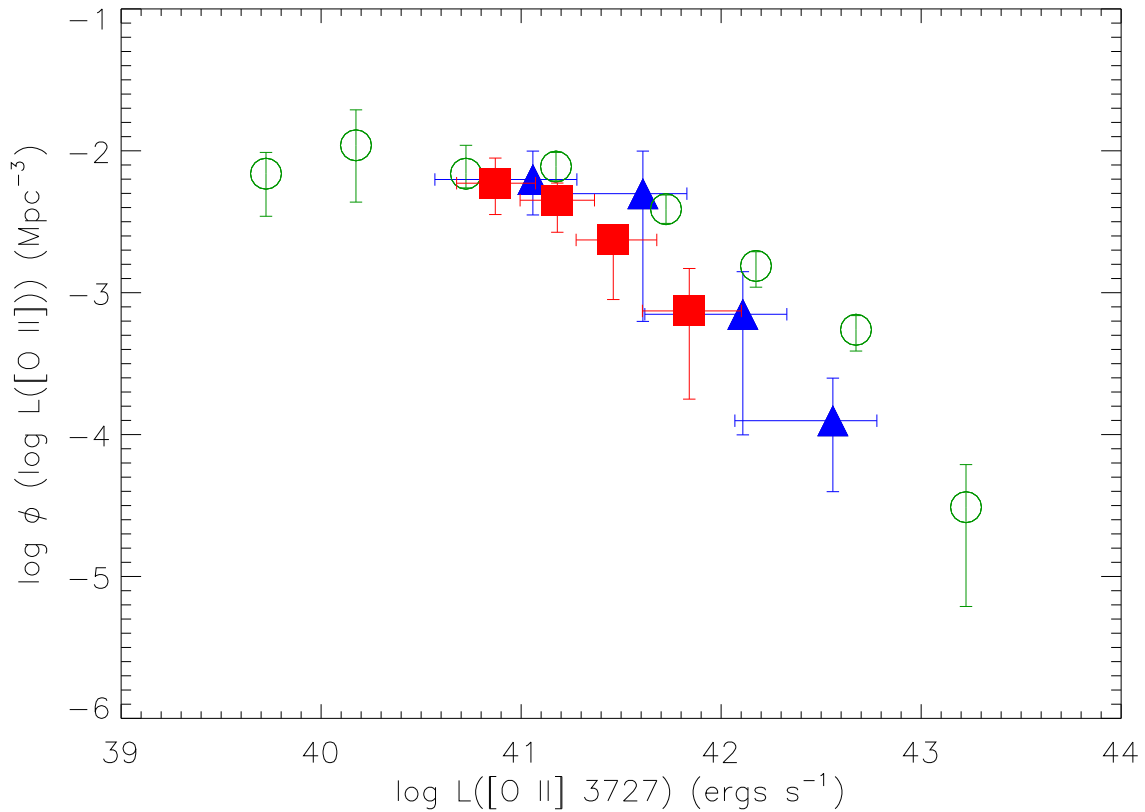}
\caption[]{(a) \ha luminosity function as found by Yan et al. (solid blue triangles;
1999), which is based on the M99 sample, transformed into an [O II]
luminosity function.  The transformed \ha luminosity function of Hopkins
et al. (solid red squares; 2000) is also included.  Both \ha luminosity
functions were transformed using [O II]/\ha=0.18, the average ratio
observed in our subset of galaxies from the total M99 sample. Open green circles
are the [O II] luminosity function published by H98.  (b) \ha luminosity
functions as found by Yan et al. and Hopkins et al., and the [O II]
luminosity function found by H98 and transformed into a \ha luminosity
function.  The transformation of the H98 luminosity function was done
using [O II]/\ha values determined by our sample of galaxies (see text for
details).}
\end{figure}

\begin{figure}
\figurenum{11b}
\centering
\plotone{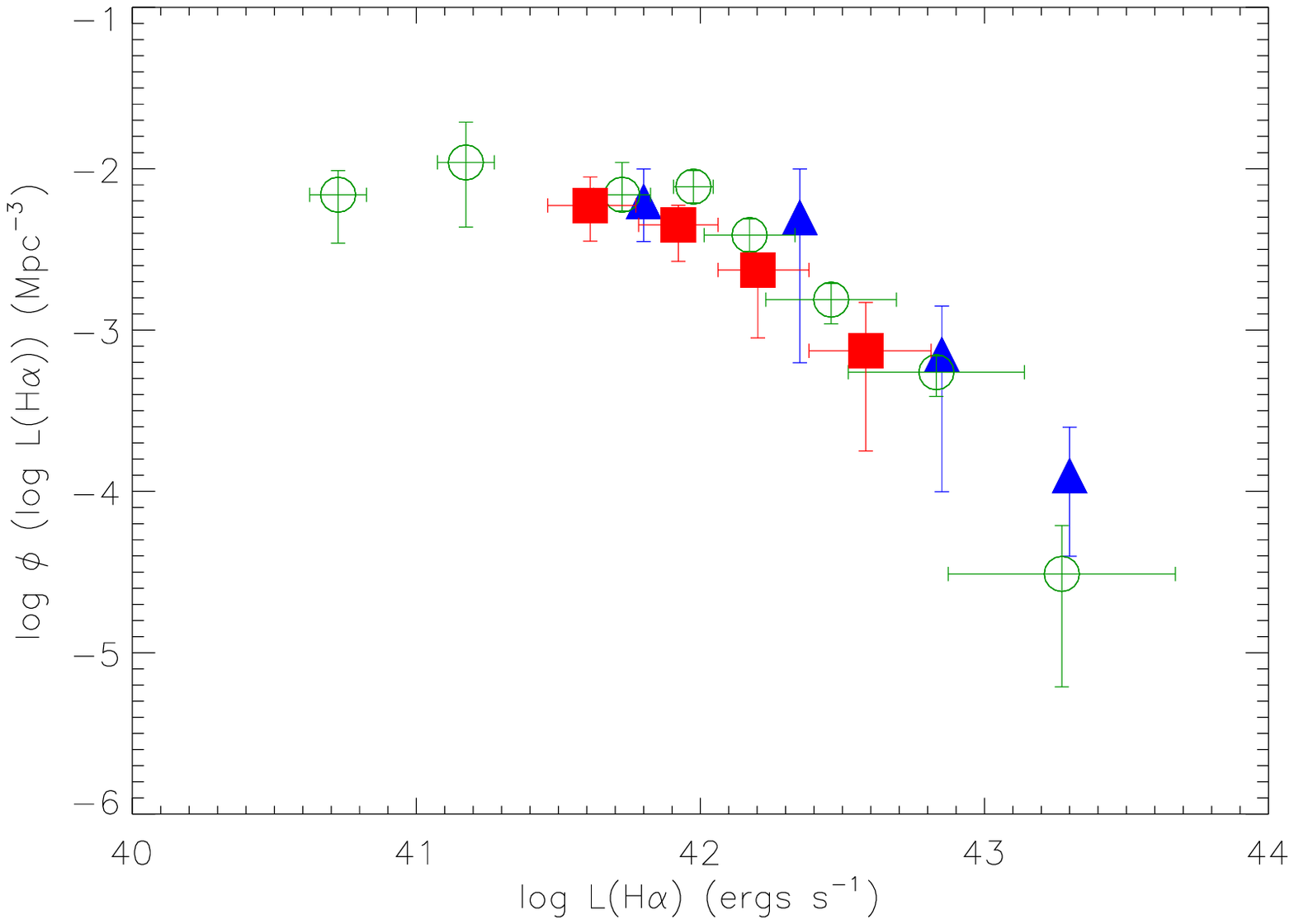}
\caption[]{}
\end{figure}

\begin{figure}
\figurenum{12}
\centering
\plotone{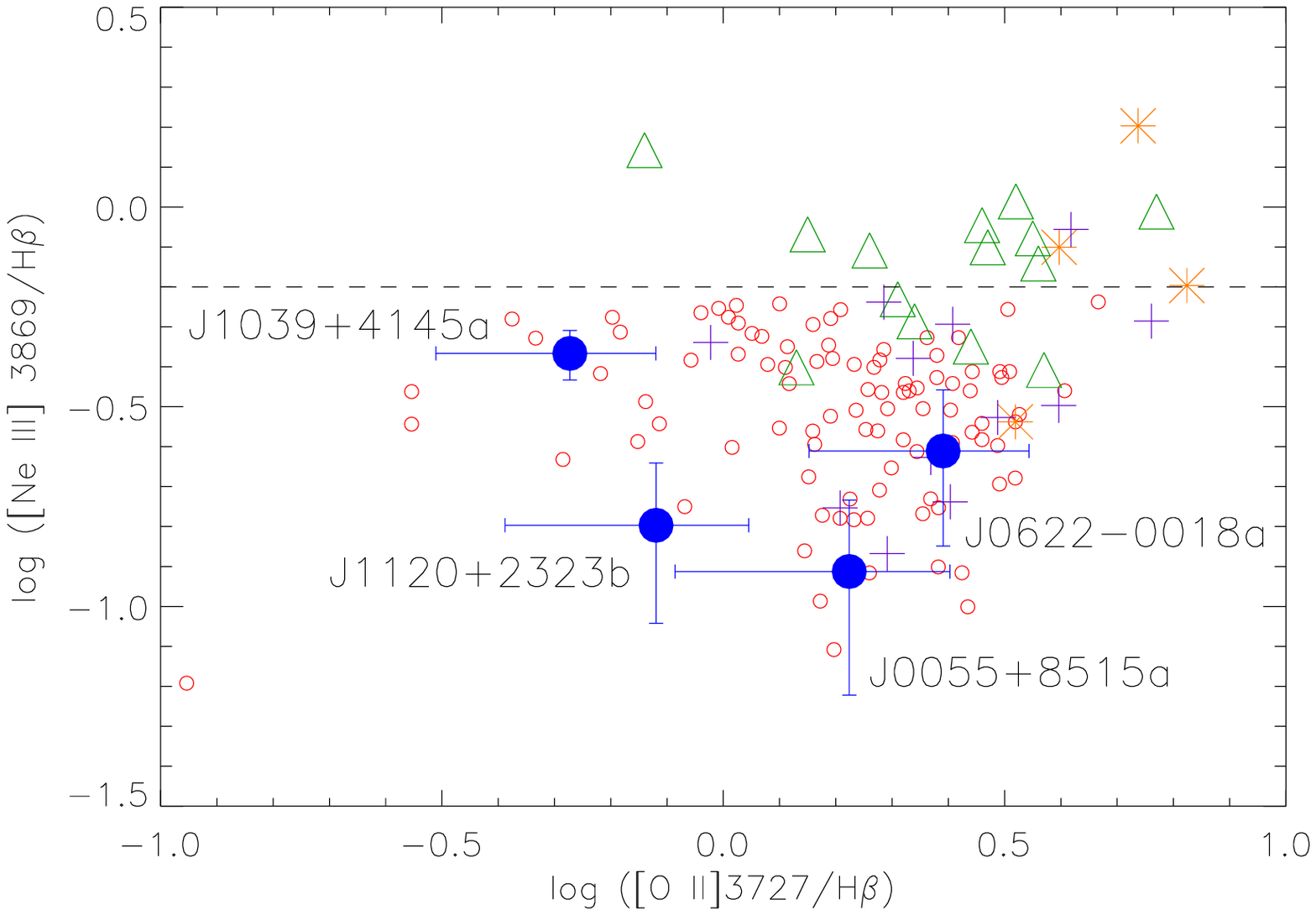}
\caption[]{[Ne III] \lam3869/\hb~vs. [O II]/\hb~intensity ratios, adapted from Rola et al. (1997).  The emission-line galaxies are classified as Seyfert 2s (green open triangles), LINERs (orange asterisks), starbursts (red dots), or ambiguous (purple plus signs).  The blue filled circles are galaxies from this study for which both [O II] and [Ne III] \lam3869 are measured confidently.  The derivations of the positions of these galaxies on the above plot assumes an \han/\hb~= 6 and [N II]/\ha = 0.5 (see the text for an explanation of how changing these assumptions affects the positions of the galaxies in this diagram).  The dashed line is the upper limit, log([Ne III] \lam3869/\hb) $\sim$ -0.2, beyond which no starburst galaxies from Rola et al.'s sample are observed.}
\end{figure}

\begin{figure}
\figurenum{13}
\centering
\plotone{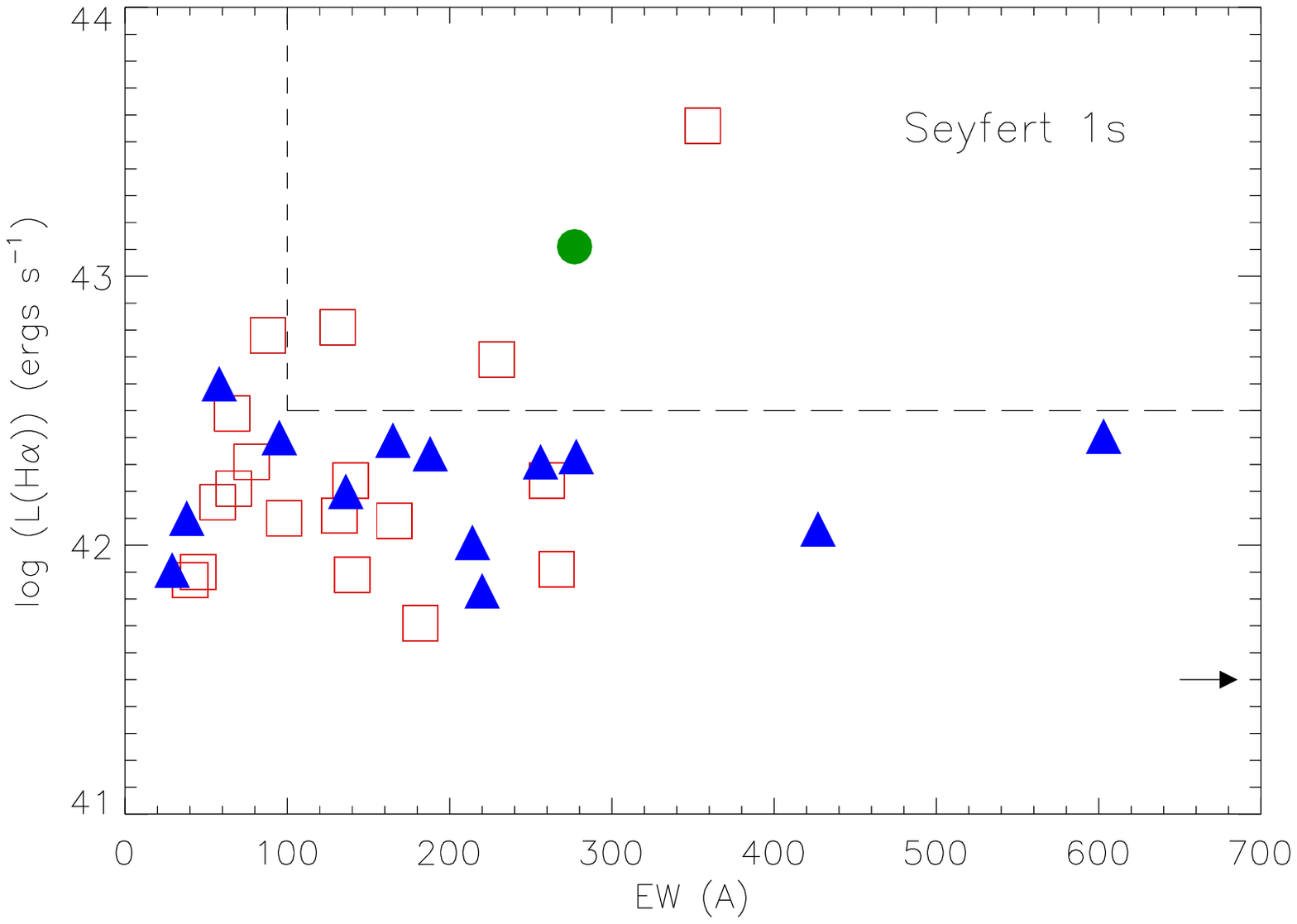}
\caption[]{Seyfert 1 diagnostic diagram using the rest-frame \ha equivalent width and luminosity.  All 33 galaxies in the M99 sample are included in the plot.  The 14 galaxies observed in our sample are represented by filled symbol: triangles for starburt galaxies and circles for the single Seyfert 1.  The remaining galaxies, those not observed in this study, are represented by the red squares.  The horizontal dashed line is the boundary $log(L_{\ha}) = 42.5$ and the vertical dashed line is the boundary EW = 100 \AA.  The region enclosed by these boundaries and labeled as ``Seyfert 1s" is where Seyfert 1s are expected to be found.  There is one galaxy that falls off to the right with an EW $>$ 700 \AA~, and has a relatively low \ha luminosity.  The galaxies that are in the Seyfert 1 region are J0917+8142a, J0917+8142c, J0923+8149a, and J0931-0449a.}
\end{figure}

\begin{figure}
\figurenum{14}
\centering
\plotone{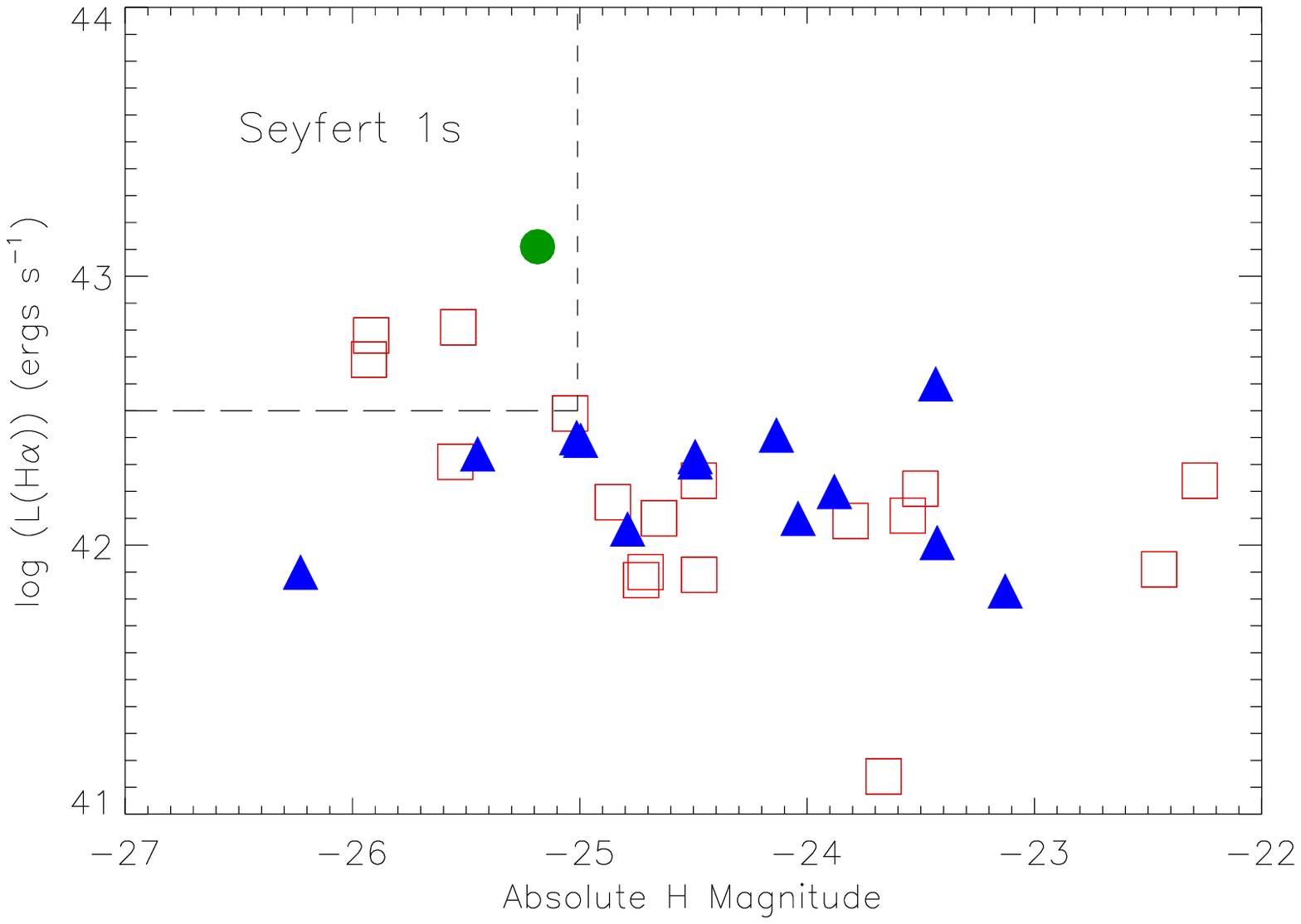}
\caption[]{Seyfert 1 diagnostic diagram using the absolute H magnitude and the luminosity of \ha emission.  All 33 galaxies in the M99 sample are included in the plot.  Symbols are as in Fig. 13.  The horizontal dashed line is the boundary $log(L_{\ha}) = 42.5$ and the vertical dashed line is the boundary $M_{H} = -25$.  The region enclosed by these boundaries and labeled as ``Seyfert 1s" is where Seyfert 1s are expected to be found.  The galaxies that are in the Seyfert 1 region are J0738+0507a, J0917+8142a, J0923+8149a, and J0931-0449a.  Two galaxies cannot be placed on the plot because of the lack of an {\it H}-band magnitude measurement, but one of them, J0917+8142c, has a high \ha flux and would most likely be placed in the Seyfert 1 region.}
\end{figure}

\clearpage

\end{document}